# SERENDIPITY IN SCIENCE[1]

**Pyung (Joe) Nahm**
Strategic Management and Entrepreneurship
Carlson School of Management
University of Minnesota
nahm0003@umn.edu

**Raviv Murciano-Goroff**
Strategy and Innovation
Questrom School of Business
Boston University
ravivmg@bu.edu

**Michael Park**
Organisational Behaviour
INSEAD
michael.park@insead.edu

**Russell J. Funk**
Strategic Management and Entrepreneurship
Carlson School of Management
University of Minnesota
rfunk@umn.edu

[1] We thank the National Science Foundation for financial support of work related to this project (grants 1829168 and 1932596). We are grateful for feedback from audiences at the 2022 Academy of Management Annual Meeting in Seattle, 2022 Strategic Management Society Annual Conference in London, 2022 Israel Strategy Conference in Herzliya, and the Strategic Management and Entrepreneurship department seminar at the Carlson School of Management at the University of Minnesota. An earlier version of this paper was the winner of the Best Paper awards in Knowledge and Innovation Interest Group at the Strategic Management Society Annual Conference and the Best Entrepreneurship and Innovation Paper awards at the Israel Strategy Conference. A portion of the computational work reported on in this paper was performed on the Shared Computing Cluster, which is administered by Boston University's Research Computing Services. All errors are our own.

**INTRODUCTION**

Serendipity, which has been defined as "the interactive outcome of unique and contingent 'mixes' of insight coupled with chance" (Fine and Deegan 1996, p. 434), is widely understood to play a central role in scientific discovery (Canon 1940; Andel 1994; Simonton 2004; Merton and Barber 2011; Fink et al. 2017). While scientists devote significant resources to planning all phases of the research process, many of the most important breakthroughs have been the result of "happy accidents." Perhaps most famously, Alexander Fleming discovered penicillin following the contamination of a petri dish in his laboratory (Roberts 1989). Making a discovery based on serendipitously encountered information—what we call a "serendipitous discovery"—requires more than just exposure to the unexpected; it also requires the ability to apprehend its potential scientific value, a sentiment expressed most famously by Louis Pasteur, who observed that "chance favors only prepared minds."

While most scholars would agree with Pasteur on the importance of a "prepared mind" for serendipitous discovery, there is far less consensus on the meaning of "prepared." Some argue that a prepared mind is an *open* one, free from preconceived notions and commitments (Barber and Fox 1958; Baumeister 2006). Others argue that a prepared mind is an *experienced* one, possessing deep knowledge in a particular domain (Merton 1973; Roberts 1989). To date, the literature has been unable to adjudicate between these two views because most research on serendipity in science is conceptual or case-study based (e.g., Barber and Fox 1958; Roberts 1989; Fine and Deegan 1996; Baumeister 2006; De Rond 2014; Yaqub 2018). Having a more complete understanding of the factors that enable scientists to benefit from chance encounters with unsought information is important, not only for advancing theoretical insight on scientific discovery, but also for developing better knowledge of whether and how serendipity might be fostered through policy action and organizational and information system design (Bawden 1986; Andre et al. 2009; Thudt et al. 2012; Makri et al. 2014; Chai and Freeman 2019; Lane et al. 2021).

In this study, we overcome the limitations of previous work and add to the ongoing debate by conducting an empirical investigation of serendipitous discovery in science that leverages a large-scale natural experiment in which scientists were exposed to unfamiliar journals.[2] Specifically, in response to the release of a new library cataloging standard known as the Second Edition of Anglo-American Cataloging Rules (AACR2), the shelving locations of scientific journals were exogenously shifted across hundreds of North American research libraries in a temporally-staggered manner (i.e., over several years). Because this policy was implemented in the 1980s, an era when scholars relied primarily on research libraries for literature search, the change in journal shelving order increased the likelihood a scientist would be exposed to unsought, but potentially useful, scientific information. Our primary objective is to empirically examine which scientists were most stimulated by the implementation of AACR2 to use unsought, serendipitously discovered journals. Using data on the 2,398,026 publications published by 523,511 scientists at 115 major public and private universities across Canada and the United States that implemented AACR2, we examine the citation behavior and innovativeness of those publications before and after the implementation of AACR2-associated shelving changes at their respective universities.

Several noteworthy findings around the "openness" or "experience" debate emerge from our analyses. First, we find that not all scientists benefit from unplanned exposure to new scientific information following AACR2 implementation. Scientists with greater *openness* were more likely to incorporate less familiar and newer references in their papers during the post-AACR2 period. Second, we find that scientists with greater depth of *experience* tended to rely more heavily on familiar and older references following the implementation of AACR2. This is likely because AACR2 created a small but impactful friction for researchers attempting to locate the journals on their library shelves. Therefore, researchers with deep experience instead relied on information that had previously been retrieved, since in that case, such friction would not be a factor. Finally, we document that the changes from AACR2 and its impact on the citation patterns of university researchers translated into changes in the innovativeness of those scientists' papers. Scientists with greater openness published more innovative papers after AACR2 than less open scientists. By contrast, there is no significant difference in the innovativeness of papers published by scientists with a greater or lesser depth of experience following the policy change. Ultimately, scientists with a more open mind appear better prepared to incorporate novel information into their research, and subsequently, the innovativeness of their research following AACR2 increased.

---

[2] We use "scientists" for expository purposes, but our data includes a broad sample of scholars, not all of whom would identify as scientists.

**THEORETICAL BACKGROUND**

**Serendipity and Science**

Serendipitous discoveries occur when a scientist utilizes serendipitously encountered information for the creation of scientific discoveries. They have long been of interest to historians, philosophers, and sociologists of science. Historical accounts celebrate how Alexander Fleming discovered penicillin based on a fortuitous accident (Lenox 1985; Cunha et al. 2010), Pasteur unearthed molecular chirality by chance (Vantomme and Crassous 2021), and Luigi Galvanti set the stage for the invention of the battery upon seeing the leg of a frog twitch during a dissection with an iron scalpel (Roberts 1989).

As illustrated in these well-known discoveries, serendipitous encounters are events or information that a scientist is exposed to that they did not specifically seek out and they are often important ingredients in scientific discoveries. When chance encounters occur, scientists are stimulated to make connections between what they experience and their own prior knowledge (Makri and Blandford 2012). Consequently, such encounters trigger scientists to think critically, creatively, and differently about their research. For example, an unexpected encounter can force scientists to reflect on their previous assumptions or ideas. Unexpected information may also inspire scientists to create connections between previous knowledge and the new information. By facilitating new connections, encounters with previously unfamiliar or unsought ideas, information, and phenomena can ultimately lead to unexpected solutions (McCay-Peet and Toms 2015).

Recognizing the important role that chance encounters play in scientific discovery, serendipity has also received significant attention from research in library and information sciences (e.g., Grose and Line 1968; Ford 1999). While researchers in these fields have generally focused on developing tools that allow users to search and locate materials efficiently (e.g., Bates 1979; Wilson, Schraefel and White 2009; Kleiner, Rädle, and Reiterer 2013), a growing stream of work has shown that when scientists browse through library resources, they are more likely to unintentionally discover unfamiliar or unsought books and journals, which may turn out to be useful (e.g., Erdelez 2005; Hinze et al. 2012; Saarinen and Vakkari 2013). In particular, physical library shelves are regular sites of serendipitous encounters (Waugh, McKay and Makri 2017). Although library users generally begin browsing through library shelves with a goal-driven search for materials, they often come across books and journals that are not relevant to their initial search goal but are proximately located to their material of interest (Toms 2000; Foster and Ford 2003). Some library users avoid using online materials for the very reason of serendipitously encountering unsought information on library shelves (Hinze et al. 2012), while libraries and information scientists have also tried to recreate library bookshelves in digital library collections to foster serendipitous discoveries (e.g., Thudt, Hinrichs, and Carpendale 2012).

**Determinants of a Prepared Mind**

Not all scientists who encounter serendipitous information, however, will benefit from it. Based on case studies and anecdotes, past work has argued that individual attributes—such as active learning, exploration (McCay-Peet and Toms 2010), and the purpose for engaging in search (Rubin et al. 2011)—can all play a role in influencing serendipity. More specifically, previous investigations of serendipity hypothesize that some scientists have minds that are more "prepared" to absorb and benefit from serendipitous information than others (Barber and Fox 1958; Lenox 1985; Lawley and Tompkins 2008; Cunha et al. 2010; Rubin et al. 2011; Makri and Blandford 2012). The prepared mind primes individuals to recognize the potential value of an unexpected event and helps them make connections and follow up on what they discovered.

Nevertheless, as noted above, there is little scholarly consensus on what precisely constitutes a prepared mind. Indeed, in their study of the history of the concept of serendipity, Merton and Barber (2011, p259) likened the notion of the "prepared mind" to a "psychological black box." Descriptions of what constitutes a "prepared mind" range from a scientist's openness (Barber and Fox 1958), traits such as alertness, flexibility, courage, and assiduity (Cunha et al. 2010), background knowledge and experience (Fine and Deegan 1996; Rubin et al. 2011; Makri and Blandford 2012), and insight (Lawley and Tompkins 2008). The most prominent conceptions of what constitutes a "prepared mind" largely fall into two categories: (1) those that emphasize the role of a scientist's *openness* and (2) those that emphasize the importance of a scientist's *depth of experience*.

The first group of research emphasizes openness as an important trait for scientists to extract value from unsought information. Openness is described as "temporary states of unfocused attention in which attention was not directed and behavior was more exploratory, more open" (Sun et al. 2011; McCay-Peet and Toms 2015). There are several reasons why openness may be essential. A scientist needs to be open to ideas that are perhaps inconsistent with his or her current views for an encounter to lead to a serendipitous discovery. Scientists who are willing to consider nonconforming information are more likely to act on that information to develop useful insights. In contrast, those who only look for new ideas that fit their existing beliefs are less likely to apprehend the potential value of a chance encounter. For scientists who only seek ideas that fit with their existing views, information that does not seem immediately relevant will be dismissed before having the opportunity to blossom (Lenox 1985). More open scientists are also suggested to be more likely to explore materials that are only loosely related to their initial search goals (Foster and Ford 2003). This exploration allows scientists to engage in a more in-depth assessment of the potential value of tangential information that they encounter. Therefore, scientists who are more open are more likely to be able to extract the latent value of new information, thereby increasing their likelihood of making serendipitous discoveries.

A second group of studies emphasizes the importance of scientists' depth of experience. According to this group, the value of unsought information is difficult to apprehend without reference to pre-existing foundational knowledge. Scholars that emphasize the depth of experience, therefore, define the prepared mind as having a detailed understanding of the topic of inquiry (Fine and Deegan 1996; Rubin et al. 2011). Depth of experience allows scientists to make connections from the new information to potential applications within their domain of knowledge. Individuals' knowledge and experience need to be connected to serendipitously encountered ideas, information, and phenomena in order to facilitate the process of scientific discovery (McCay-Peets and Toms 2015). Inexperienced scientists may not recognize the usefulness of data, phenomena, or contacts that they encounter by chance because they lack a rich catalog of contextual knowledge. Without extensive experience in a particular subject area, scientists would have difficulties drawing connections between the domain and unsought knowledge to facilitate meaningful discoveries and may simply consider unsought knowledge as tangential. Therefore, depth of experience represents a critical tool for scientists to unlock the potential value residing in unsought information.

In summary, one camp suggests that the openness of a scientist is critical for allowing unplanned encounters to stimulate serendipitous discoveries. The other camp argues that serendipitous discovery requires the scientist to recognize the significance of new information relative to what is already known and therefore emphasizes the importance of depth of knowledge in a particular domain. In what follows, we address this tension by using the implementation of a new library cataloging policy, AACR2, as a natural experiment to develop a more complete understanding of the prepared minds' dual dimensions. For an unforeseen scientific insight to transpire in the library, scientists must first have the chance to come across an unfamiliar journal. In other words, serendipitous discoveries entail the exposure to unsought knowledge before the prepared mind of a scientist can start to play a role. By exogenously shifting the locations of scientific journals, AACR2 provides an opportunity for the prepared mind of a scientist to potentially be open to serendipitous discoveries. Therefore, leveraging this natural experiment, we can examine how openness and depth of experience affect scientists' use of new knowledge and information and their ability to benefit from serendipity.

**EMPIRICAL CONTEXT**

In 1967, the Library of Congress, American Library Association, Library Association of the United Kingdom, and Canadian Library Association published the first edition of the Anglo-American Cataloging Rules (AACR1), which aimed to establish systematic and universal rules for organizing library materials (Martell 1981; Joudrey, Taylor, and Miller 2015). While the AACR1 guidelines were comprehensive and offered rules for all types of materials found in library collections, for our purposes, the most important were those addressing the shelving order of journals (Hopkins and Edens 1986). Under the AACR1 system, journals were generally placed on library shelves alphabetically according to their title. Thus, the *American Journal of Physics* and the *American Journal of Political Science* would be shelved in close proximity, even though they are topically very different. This alphabetical approach differs from the perhaps more familiar ones used for the placement of books, which are usually topic-based (e.g., Dewey Decimal classification). Importantly, AACR1 included a critical exception to the general rule of alphabetical shelving by title, which applied to journals with titles that included the name of a corporate author (e.g., the American Medical Association). For these journals, shelving placement was determined alphabetically, but by the name of the corporate author, not the title of the journal (Gorman, 1981; Hopkins and Edens, 1986).

In 1974, representatives from national libraries and library associations updated the AACR1 standard, and in 1978, released a second edition known as AACR2, before officially adopting the new standards in 1981. Among the changes was a revision to the rules for cataloging journals, which dropped the previous exception for publications that included the name of a corporate author in their title. Consequently, journals that were placed on library shelves alphabetically by the name of a corporate author under AACR1 required re-shelving under AACR2 (Gorman 1981).[3] As an example, under AACR1, the *Journal of American Medical Association* (JAMA) would be shelved alphabetically as though its title were, "American Medical Association. Journal." After AACR2 was implemented, JAMA's placement would be determined by its actual title, "Journal of American Medical Association." As a result, pre- and post-AACR2, the shelving location of JAMA changed from being proximate to journals with titles beginning with "A" (e.g., *American Malacological Bulletin, American Midland Naturalist*) to journals starting with "J" (e.g., *Journal of Acquired Immune Deficiency Syndromes, Journal of the American Mathematical Society*).

---

[3] For journals with titles that did not include the name of a corporate author, no changes in placement were made.

At the time of AACR2's implementation, academic scientists relied on physically visiting university libraries and locating journals on library shelves to seek research-relevant information. While library shelves assisted scientists in achieving their information-seeking goals (Waugh, McKay and Makri 2017), library shelves also form a type of serendipity engine by exposing users to information that they had not actively searched for when browsing the shelves (e.g., Erdelez 2005; Hinze et al. 2012; Saarinen and Vakkari 2013). As the change from AACR1 to AACR2 shifted the shelving orders of journals, the likelihood of academic scientists encountering new information increased, both through impacting where scientists searched for familiar (to them) journals and the (presumably unfamiliar) journals in proximity to them. As an illustration, consider a scientist that reads and uses articles published in *JAMA*. After AACR2 implementation, the scientist finds *JAMA* in a new shelf location in the library, proximate to a set of journals that were previously not proximate to *JAMA*, thus becoming exposed to new and unsought journals.

For our study, several considerations make AACR2 a valuable empirical context. First, prior literature has established that shelving rules are critical determinants of information exposure in libraries, governing everything from the amount of time patrons will spend searching (based on the efficiency of the system) and the physical space they will traverse (based on its layout) to the specific materials (both sought and unsought) they will encounter during their visit (e.g., Hancock-Beaulieu 1993; Loose 1993; Svenonius 2000; Kleiner, Rädle, and Reiterer 2013). Consequently, the shift from AACR1 to AACR2 offers a unique window into the relationship between information exposure and scientific discovery. The nature of the transition is also fortuitous, as journals were re-shelved based on their title, not on content, discipline, or other factors that are likely most pertinent for scientific search. Indeed, from this perspective, the implementation of AACR2 is reminiscent of the serendipity-maximizing system envisioned by Grose and Line (1968), according to which materials would be placed on library shelves in random order.

Second, AACR2 is valuable from a study design perspective, as there are natural control groups at both the journal and university levels. As noted above, the revised guidelines introduced by AACR2 only applied to publications with corporate author names included in their titles which, prior to 1981, accounted for approximately 16.5% of journals in the Library of Congress catalog. Consequently, we are able to categorically distinguish journals that were treated from those that were not. In addition, while many universities adopted AACR2 in 1981, the rollout was temporally staggered, with some implementing the new rules as early as 1979 and others waiting as late as 1989, which allows us to separate universities into treatment and control groups.[4,5]

Finally, AACR2 is an attractive setting because the changes it entails are exogenous to our outcomes of interest. Specifically, we suspect that scientists are not likely to base their decision to read and use the articles from a particular journal in their work on whether or not the journal includes a corporate author in its name, but instead are likely to focus on considerations like content or reputation, neither of which were factors in the AACR2 shelving changes. In support of this assumption, in Appendix A, we examined the average number of citations received by journals prior to 1981 and found no systematic evidence that journals with corporate authors in their names were cited more than journals without corporate authors in their names.

# DATA

## 2.1 Data and Sample

The raw data used for analysis in this paper derives from merging information in two bibliometric databases along with hand-collected information about the implementation of AACR2 at universities in

---

[4] AACR2 was officially adopted in 1981 by national libraries of the U.S. and Canada, the Library of Congress and the National Library of Canada (now Library and Archives Canada), but some university libraries implemented AACR2 before 1981 while others adopted it after 1981. University libraries adopted AACR2 at different times for a variety of reasons. First, national libraries postponed the implementation of AACR2 twice, once in 1978 and once in 1980, before finally implementing in 1981. Because university libraries anticipated national libraries to eventually adopt AACR2, some libraries made the change before the official adoption year. Second, our semi-structured interviews with librarians revealed that the reshelving of journals is an arduous process and therefore, libraries with fewer resources adopted AACR2 in later periods.

[5] Our difference-in-differences setup leverages variations in the timing of AACR2 implementation across universities. Hence, the control groups (i.e., researchers at universities that have not yet implemented the change) do eventually get treated when their universities implement AACR2. This setup of difference-in-differences is similar to ones used in prior research (e.g., Kim 2022)

North America. The first source of data comes from the *Web of Science* (WoS) database, which contains bibliometric information on scientific publications. The WoS data have been widely used in prior, related research (Evans 2008; Rawlings et al. 2015) and are generally thought to be representative of publication and citation activity across most major fields of science and scholarship (Birkle et al. 2020). An observation in this database is a published scientific work. Each of these records includes metadata about the work, including the authors, the university affiliations of each author, and the articles cited in the reference list of the work.

The second source of data is the Library of Congress's catalog. Using this data source, we identify the set of journals that changed shelf position as a result of AACR2. The Library of Congress's Open-Access Distribution Service data contains information on nearly all journals ("serials") published over time. This catalog provides the title of each journal, an identifier for the journal (ISSN), and information on whether the journal had a corporate author. From this information, we flagged journals as either having the corporate author's name in the title—and thus being affected by AACR2—or not.[6] Of the 9,750 English-language journals in the Library of Congress records, we identified 601 as having corporate author names in their title. In addition, the Library of Congress records also enable us to measure how far apart journals would have been positioned on library shelves for libraries.[7] We refer to this measure as the shelf-distance between two journals.

With this data in hand, we turned our focus to identifying the years in which universities in the U.S. and Canada implemented the AACR2 standard in their research libraries. We chose our sample universities from members of the Association of Research Libraries (ARL). ARL is a nonprofit organization of 125 research libraries of major public and private universities, federal government agencies, and large public institutions in Canada and the U.S. Research on libraries and library cataloging have frequently used ARL member libraries as the representative sample of research libraries (e.g., Case and Randall 1992; Russell 2004; Heady et al. 2021). Starting with the 115 member libraries of the ARL located at major public and private universities, we searched library websites, library journals and bulletins, and historic college newspapers for information about when each library implemented the AACR2 shelving change. For libraries without information available in those sources, we emailed the librarians and directors of those libraries, asking if they had records of when the library implemented the change. Out of 115 libraries, we obtained information on the exact year of AACR2 adoption for 49 libraries. In that sample, 31 libraries implemented AACR2 in 1981 (the year that AACR2 was enacted), while 18 libraries implemented it before or after 1981. In our analyses, we impute 1981 as the implementation year for universities for which we do not have an exact implementation year, but our results are robust when we only use universities with information on the exact date of AACR2 implementation.[8]

---

[6] Appendix B provides more details on how we obtained data on corporate author names of journals and how journals are coded as having corporate author names in their title.

[7] We measure this as the number of journals that appear in the catalog between two journals given the AACR and AACR2 cataloging rules. This measure assumes that a library subscribed and shelved all the serials present in the Library of Congress catalog. That being said, the measure can still be used as a relative distance and provide the upper-bound on the possible distance between two journals for libraries that subscribed or shelved fewer journals than the full Library of Congress collection.

Finally, we need to place scientists at institutions during specific points in time to see which scientists may have been affected by AACR2 in a particular year. We choose to use the corresponding author's affiliation because the corresponding author, compared to other individuals listed as authors on a paper, is more likely to be responsible for leading the project (Wren et al. 2007, Mattsson et al. 2011; Bhandari et al. 2014). Additionally, WoS begins to have reliable affiliation data for all authors, including authors that are not corresponding authors, starting in 2008. Consequently, due to data limitations, considering the affiliation of all authors within our sample would not have been feasible.

---

[8] In Appendix C, we run analyses without universities that we imputed 1981 as the AACR2 implementation year. This leaves us with 45 universities. Across all models in Appendix C, results are consistent with our analyses with universities with imputed AACR2 years. These results add confidence that the subset of universities with imputed AACR2 implementation years do not systematically change the relationships that we observe in our analyses.

## 2.1 Three Levels of Datasets

From these data sources, we construct three datasets. First, to demonstrate how scientists changed their use of journals, we built a shelf-position citation dataset. This dataset is created by finding all of the researchers who cited a journal in 1977 that would eventually be moved during the implementation of AACR2. We call these previously cited journals the "treated" journals. For each of these researchers and each of the treated journals they cited in 1977, we find the journals shelved within 10 journals away on either side of the journal on the library shelves in 1977, before AACR2, as well as in 1983, after AACR2 was implemented. We call those journals the "proximate" journals. Finally, we attach the probability that the researcher had cited each of those proximate journals in 1977 as well as in 1983. Using this data, we document how moving a journal that a researcher previously used also changed the probability that a researcher cited proximate journals.

  Second, to demonstrate the dynamics of the effect of AACR2 on scientists' citation behavior, we construct the scientist-panel. This dataset is created using the WoS data on the corresponding authors and their associated publications. An observation in this dataset is a scientist × university × year. For each observation, we note the journals that scientists cited in their publications, and we flag whether these journals were within the set of journals that moved because of AACR2. We also record whether the scientist's affiliated university library had implemented AACR2 in the year of observation.

  Finally, for our analyses that demonstrate the effect of AACR2 on both scientists' citation behavior and publication outcomes, we construct a dataset at the publication-level. This dataset is a repeated cross-section of publications that is constructed with multi-level variables. We have university-level data on the AACR2 implementation year. We have data on corresponding authors of publications, and we also have data on individual publications. Our sample consists of 2,398,026 observations from 1978 to 1997. With this dataset, we investigate which scientists take advantage of serendipitous exposure to new knowledge and produce innovative publications.

# 3. ANALYSES

## 3.1 Serendipitous Encounters and the Use of Proximate Journals

We begin by demonstrating that the change in the shelving order of journals due to AACR2 increased opportunities for scientists to serendipitously encounter scientific journals. If a scientist previously used a journal and that journal moved to a different location, then we would expect that the scientist would be less likely to use proximate journals from the previous shelving location. Similarly, we would expect that the scientist would be more likely to encounter and use the proximate journals in the new location.

Using the shelf-position citation dataset, we analyze how the propensity to cite journals proximate to treated journals changed from before AACR2 to after AACR2 was implemented. First, we examine how scientists changed their propensity to cite journals proximate to the location where a treated journal was *removed from* following AACR2. Figure 1(a) shows the average differences in the probability of citing journals between 1977 and 1983 based on their shelf distance to the moved journal. The plot demonstrates that the probability of citing journals closest to the former position of the moved journal declined in the years following the focal journal being moved. The decrease in the probability of citing a journal is on average larger for the journals closest to the previous position of the focal journal.

------------------------------------
Insert Figure 1 here
------------------------------------

Second, we examine how scientists changed their propensity to cite journals proximate to the location where a journal was *moved to* following AACR2. Figure 1(b) shows the average differences in the probability of citing journals between 1983 and 1977 based on their shelf distance to the moved journal's new position. The probability of citing a journal in that new location increased for all proximate journals. The increase is most pronounced for the journals nearest to the new location of the focal journal.

These plots demonstrate that by changing the shelf position of journals, AACR2 created opportunities for scientists to encounter and use different sets of journals. By changing which journals were in proximity to each other, AACR2 increased the probability that scientists would be exposed to different sets of journals than they may have previously seen.

## 3.2 Utilization of Unfamiliar and Familiar Journals

Given that AACR2 pushed scientists to encounter different sets of journals on library shelves, did scientists simply increase their use of journals that they had previously used in their research (familiar journals), or did these encounters also inspire scientists to pick up and cite journals that they had not previously used (unfamiliar journals)? To answer this question, we trace out the dynamic effect of AACR2 by estimating an event study model with scientist-panel data. The model is based on the following specification:

$$(1)\ OUTCOME_{it} = \sum_{j=-4}^{10} \gamma_j D_{m,t}^{j} + \mu_f + \mu_m + \psi_{s,t} + \epsilon_{i,t}$$

In the above, the $D_{u,t}^{j}$ is a set of dummy variables, which takes on the value of 1 when the adoption of AACR2 at university $u$ occurred $j$ years in the future. The coefficients of interest are the $\gamma_j$, which tells us the effect of AACR2 on outcomes for the scientists in each time period. Besides tracing the dynamic effects of AACR2 going forward, if the coefficients for the time periods before the adoption of AACR2 are significantly different from zero, this could imply a violation of the parallel-trends assumption and the possibility that scientists anticipated the adoption of AACR2.

Figure 2 estimates the event study regression specified in Equation (1) in order to measure how scientists' citations of journals changed over time. In Figure 2(a), we plot the estimated coefficients on the event study indicators when the dependent variable is citations to unfamiliar journals. In this plot, the estimated coefficients are positive and significant immediately following the implementation of AACR2 at their university. The coefficients range between 0.01 and 0.05, implying that scientists were between 1 and 5 probability points more likely to cite an unfamiliar journal after AACR2 than in the year just before AACR2 was adopted. Relative to the unconditional probability of citing an unfamiliar journal, 0.693, this amounts to between 1.4%-7.2% increase in citations to these journals. These coefficients demonstrate that the serendipitous encounters that scientists had with proximate journals that were unfamiliar to them increased their likelihood of reading and citing those journals in subsequent years. In contrast, Figure 2(b) plots citations to familiar journals. The estimated coefficients in that plot are close to zero and not statistically significant in the years following AACR2. In other words, after encountering a new set of journals, scientists increased their use of such journals—incorporating new sources of information—but did not change their use of journals that they had previously encountered.

-----------------------------------
Insert Figure 2 here
-----------------------------------

### 3.3.1 Scientists and AACR2 Implementation

Our primary objective in this study is to adjudicate whether the "prepared mind" is based on a scientist's openness or depth of experience. In this section, we empirically investigate which scientists were most likely to change their citation patterns because of the implementation of AACR2. Given that unfamiliar journals are more likely to be cited after AACR2 implementation, we examine how scientists' openness and depth of experience affect scientists' likelihood of using unfamiliar journals post-AACR2. We also explore whether scientists' use of moved and familiar journals changes conditional on the scientist's openness and depth of experience. Lastly, we look into how AACR2 implementation affects the innovativeness of scientific works produced by scientists.

**Measurement of Key Variables.** To investigate how scientists' utilization of scientific journals changed in response to the implementation of AACR2, we examine citations made by scientists in publications. To measure changes in citation behavior, we define three binary indicators: *cites a moved journal, cites an unfamiliar journal,* and *cites a familiar journal*. The variable *Cites a moved journal* takes a value of 1 if the publication cites works from journals that have corporate names included in their title and therefore moved to a different shelf following AACR2, and 0 if otherwise. *Cites an unfamiliar journal* takes a value of 1 if the publication cites works from journals that the corresponding author has not cited before, and 0 if otherwise. Lastly, *Cites to a familiar journal* takes a value of 1 if the publication cites works from journals that the corresponding author has cited in the past, and 0 if otherwise.

To measure the impact of AACR2 on the scientific works produced by scientists, we attach a measure of innovativeness to each publication produced by a scientist in our sample. Developed by Funk and Owen-Smith (2017), the *CD index* computes publication innovativeness based on the network pattern of citations to a publication. The *CD index* measure ranges from -1 to 1. Scores less than 0 imply that the publication reinforces or consolidates existing streams of literature by drawing attention to prior work; scores larger than 0 mean that the publication destabilizes or disrupts existing streams of literature by drawing attention away from prior work. We measure publication innovativeness using a binary version of the CD index. We define an innovative publication as a *Destabilizing publication* for papers with CD indexes above 0. We define a less innovative publication as a *Consolidating publication* for those papers with CD indexes below 0.

Recall that our main explanatory variable is the implementation of AACR2 in university libraries. *Post-AACR2* takes a value of 1 in years in which the corresponding author's university library implemented AACR2 changes, including the year the university adopted AACR2, and 0 if otherwise.

We examine how the impact of AACR2 on a scientist's publication outcomes and citation behavior differs by the "prepared mind" of the scientist. Prior research describes the prepared mind of the scientist with both openness and depth of experience. Openness is often measured with the breadth of knowledge that was used (e.g., Laursen and Salter 2006; Tsinopoulos, Yan and Sousa 2019). Following prior research, we measure *openness* as the proportion of the corresponding author's backward citations to the corresponding author's non-core fields prior to the year of publication, where the core field is defined as the WoS subject area most cited by the corresponding author. If a scientist cites more works from fields that are not his/her core area, then this implies that the scientist is more receptive and interested in ideas and concepts that are used in other subject areas. Hence, we believe that the proportion of citations to non-core areas is a good proxy for the openness of the scientist.

We measure the *depth of experience* with the corresponding author's publications to the core subject area from the beginning of their career up to the year prior to the focal year. Depth of experience is measured as the proportion of the corresponding author's past publications in the corresponding author's core field (Boh, Evaristo, and Ouderkirk 2014), where the core field is defined as the WoS subject area most published in the corresponding author. Scientists that publish more in their core area are more likely to have deeper knowledge in that area than others who publish in diverse areas after controlling for total publications. Hence, the greater the proportion of publications in the core field, the higher the depth of experience of the scientist.[9]

For the purpose of precision and addressing alternative mechanisms affecting a scientist's citation behavior and publication outcomes, we included several control variables in our publication-level models. We control for *Ln(total number of publications)*, which is the natural log of the total number of publications published by the corresponding author prior to year *t*. This variable helps to account for heterogeneity in publishing experiences and capabilities of the corresponding author. We also control for *Ln(number of references cited)*, which is the natural log of the total number of citations made by the focal publication. If more citations are made by the focal publication, there are higher chances that the publication will cite a variety of journals which can impact both the CD index and the type of citations made by the publication. *Ln(publication team size)* accounts for the number of co-authors for the focal publication, as the size of the team can influence the quality of publication because larger teams have a more diverse pool of knowledge and wider access to a network of scientists. Lastly, we include university, subject field, and year fixed effect. Table 1 shows descriptive statistics for the variables used in this next set of analyses.[10]

---
Insert Table 1 here
---

**Methodology.** For estimating the effect of AACR2 on innovative outcomes, we use a difference-in-difference approach with the publication-level data. Our approach is based on the staggered implementation of the AACR2 shelving order across universities over time. Specifically, we estimate the following empirical specification:

---

[9] In Appendix D we show that measures of openness and depth of experience capture distinct variation. There are significant proportion of scientist-year observations in each quadrant of high/low openness and high/low depth of experience.
[10] See Appendix E for a correlation matrix of key variables.

$$(2)\ OUTCOME_{uizt} = \beta_1 AACR2_{ut} + \beta_2 AUTHOR\_CHAR_i + \beta_3 AACR2_{ut} AUTHOR\_CHAR_i +$$
$$\Omega_{it} + \alpha_u + \alpha_z + \alpha_t + \varepsilon_{uizt}$$

In the above equation, $u$ represents a university, $i$ is a publication, $z$ is the subject field of the publication, and $t$ is a year. The variable $AACR2_{ut}$ represents an indicator that takes the value of 1 when university $u$ at time $t$ had implemented AACR2 in their library. $AUTHOR\_CHAR_i$ is a vector that contains our measure of openness as well as our measure of depth of experience of corresponding authors. The vector $\Omega_{it}$ represents time-varying controls. We also include fixed effects for the university, $\alpha_u$, the subject field, $\alpha_z$, and the year, $\alpha_t$. The coefficient of interest in the above regression is $\beta_3$, the coefficient on the interaction of AACR2 being implemented and the measures of openness and depth of experience. This coefficient tells us how the impact of AACR2 differs across scientists based on their openness and depth of experience.

**Challenges.** Our empirical strategy relies on the timing of AACR2 implementation being exogenous to the decisions of individual scientists regarding which journals to cite. We address the main potential identification concerns as well as the evidence that placates these concerns below.

First, the AACR2 shelving system may reflect scientists' desires regarding where to shelve journals. Specifically, if scientists who are more innovative and have a tendency to use more unfamiliar journals campaigned for the library association to change the AACR2 in a way that benefited their use of the library, this would be a concern. The details of the AACR2 system suggest that this is not the case. The AACR2 cataloging system switched from ordering journals by their corporate authors to using the full title. This alphabetic ordering, therefore, could not have favored highly innovative scientists versus less innovative scientists.

A second challenge will be if there is a selection in which libraries adopted AACR2. For example, if AACR2 adoption was correlated with scientists having broader or more narrow research agendas (or if AACR2 adoption was correlated with an entire university having a broader or more narrow research focus), this could confound our analysis. This concern regarding selection is unlikely to influence our results. AACR2 became the standard at university research libraries and nearly every U.S. university implemented it within a few years following its release.

A third challenge is if university libraries anticipated the creation of AACR2 and made changes before the implementation of the new shelving system. In that case, our estimates of the effect of AACR2 would likely be attenuated. We address this concern by identifying the exact year when each university library implemented AACR2. The results from event study plots, shown in Figure 2, also do not show a significant difference between the behaviors of scientists at universities that had implemented AACR2 versus those that had not in the time period before AACR2 went into effect. Furthermore, the event study evidence, based on estimating Equation (1), shows that the changes in the scientist citation behavior occur immediately when AACR2 goes into effect at their university.

### 3.3.2 The Prepared Mind: Heterogeneity in AACR2's Effect on Scientists' Use of Journals

Exposure to different sets of journals may have differential impacts on scientists with different backgrounds and characteristics. While theory suggests that scientists with a "prepared mind" are going to be the most likely to derive benefits from serendipitous encounters, it is unclear whether possessing a prepared mind is more related to an openness to diverse sources of knowledge or a depth of experience.

In Table 2, we use publication-level data to test which scientists were most impacted by the AACR2 shelving changes. First, in models M1 and M2, we estimate specification (3) with the dependent variable being *Cites a moved journal*. In M2, the estimated coefficient of the interaction term on AACR2 and the openness of the scientists is 0.031 and significant ($p$=0.000). That coefficient means that following AACR2, the scientists with the highest level of openness were 6.34% more likely to cite a journal that had moved positions than the scientists with the lowest level of openness. In other words, scientists who exhibited openness prior to AACR2 increased their use of moved journals following AACR2. While open scientists increased their use of moved journals, scientists with more depth of experience did not change their utilization of those journals.

---------------------------------
Insert Table 2 here
---------------------------------

In models M3-M6, we show byproducts of increasing use of journals that moved by examining to what extent scientists changed their propensity to cite unfamiliar and familiar journals following AACR2. Specifically, in M3, we regress an indicator for whether scientists cited a journal that was previously not cited by them (unfamiliar journals) on a difference-in-difference specification. In M4, we add an interaction between the openness of the scientist as well as the depth of experience of the scientist and the indicator for AACR2 being implemented at that scientist's university. The coefficient on the interaction term between AACR2 and openness is 0.023 and significant ($p=0.000$). This coefficient means that the scientists with the highest level of openness are 3.15% more likely to cite an unfamiliar journal than the scientists with the lowest level of openness, following AACR2 relative to those who are not open. This demonstrates that these scientists with high openness increased their use of previously unfamiliar journals. In other words, more open scientists encountered unfamiliar journals and incorporated those into their papers going forward. The interaction term with the depth of experience of a scientist is -0.016 ($p=0.023$). This coefficient says that while open scientists increased their use of unfamiliar journals in response to seeing new journals after AACR2, scientists with more depth of experience actually avoided using those new journals. Specifically, scientists with the highest level of depth of experience are 2.31% less likely to cite unfamiliar journals than scientists with the lowest level of depth of experience after AACR2.

How did open and experienced scientists respond to AACR2 in their citations to familiar journals? In models M5 and M6, the dependent variable is whether the publication cites a journal that the scientist had previously cited in one of their publications (familiar journals). The coefficient on the interaction term between AACR2 and openness is -0.061 and significant ($p=0.000$), implying that very open scientists leveraged familiar journals at a lower rate after being exposed to new sources of knowledge from AACR2. Relative to the unconditional probability of citing a familiar journal, this coefficient says that the most open scientists were 6.93% less likely to cite a familiar journal than the least open scientists after AACR2. Again, however, scientists with more depth of experience do the opposite. The coefficient on the interaction with experience concentration is 0.022 and significant ($p=0.000$), which implies that scientists with the highest level of depth of experience were 2.64% more likely to cite familiar sources than scientists with the lowest level of depth of experience following AACR2.[11]

The above results reveal contrasting effects when the location of the journals that scientists use is shuffled, and scientists are potentially exposed to new sources of knowledge. Scientists that are more open are more likely to leverage exposure to new sources of information. In addition to utilizing journals that moved, these scientists specifically gravitate towards incorporating materials from journals they had not previously used. In contrast, for scientists with a high depth of experience, whose work draws upon a specific set of works, creating friction in their access to these materials is challenging. Thus, AACR2 and the exposure to serendipitously seen scientific material primarily benefited open scientists whose previous work prepared them for incorporating novel information into their research.

### 3.3.3 Evaluating AACR2's Effect on Innovation and Scientific Research Output

How did AACR2—and the serendipitous encounters with new sets of journals—affect the innovativeness of scientists' publications? Table 3 reports estimates of the effect of AACR2 on the innovativeness of the research produced by scientists using difference-in-differences specifications and the publication level data.[12] The dependent variables in these specifications are binary variables that indicate whether the publication is a destabilizing publication (i.e., more innovative publication) or not, and whether the publication is a consolidating publication (i.e., less innovative publication) or not. Models M1 and M3 display the estimates when controlling for scientist attributes, such as the scientist's openness, depth of experience, career experience, and university, as well as publication attributes, such as the team size and subject field of the publication. The coefficients on AACR2 are not statistically significant in both models. This implies that when pooling all scientists together, on average, AACR2 did not have a discernable effect on the innovativeness of research outputs.

---

[11] We replicate Table 2 with count versions of dependent variables in Appendix F. Dependent variables are number of moved journals cited, number of familiar journals cited, and number of unfamiliar journals cited. We employ pseudo-maximum likelihood Poisson models and find similar results.

[12] A burgeoning literature addresses the challenges of analyzing the staggered roll-out of policies using difference-in-differences methods (e.g., Goodman-Bacon, 2021). The main concerns that this literature highlights occur when the effect of a policy has heterogeneity across the treated cohorts. In the setting of AACR2, however, this concern is somewhat mitigated since the vast majority (87.83%) of the universities in our sample adopted AACR2 in the year 1981 (see Appendix G for AACR2 implementation years). Therefore, while there may be heterogeneity across universities, our estimates are less likely to be influenced by differences across cohorts. Nonetheless, we run models in which early adopters do not become controls for later adopters to address these concerns. In Appendix H, we limit the sample to 1980, 1984, 1988, and 1989 treated-cohorts (excluding cohorts with imputed AACR2 years), but only use the data through the year 1983 for the 1980 cohort and through the year 1987 for the 1984 cohort. The logic is that we are excluding the time when late adopters are to be compared against the earlier adopters. We find similar results for these analyses. More open scientists are more likely to produce destabilizing publication, less likely to produce consolidating publication, and more likely to cite unfamiliar journals.

While the overall impact of AACR2 on outputs may be insignificant, subsets of scientists may have been positively affected and others negatively affected. In models M2 and M4, we include interaction terms for the implementation of AACR2 and the openness and depth of experience of scientists (i.e., corresponding authors of publications). In both models, the coefficients of AACR2 remain insignificant. In contrast, in model M2, the coefficient on the interaction between AACR2 and openness is positive and significant ($\beta = 0.033$, $p = 0.000$) while the moderation of depth of experience is insignificant ($\beta = -0.006$, $p = 0.352$). This implies that the scientists with the highest level of openness are 8.01% more likely to publish a destabilizing publication than the scientists with the lowest level of openness after AACR2 implementation. Interestingly, the interaction between AACR2 and openness in model M4 is negative and significant ($\beta = -0.023$, $p = 0.002$) while the interaction between AACR2 and depth of experience is positive and significant ($\beta = 0.013$, $p = 0.035$). These results suggest that the most open scientists are 4.07% less likely to publish a consolidating publication after AACR2, while scientists with the highest depth of experience are 2.52% more likely to publish a consolidating publication. In summary, the results suggest that not all scientists benefit from the serendipitous encounters created by AACR2. Scientists who are more open benefited in the innovativeness of their publications, while scientists with deep depth of experience were not able to increase innovativeness because they did not leverage serendipitous encounters brought by AACR2.

-----------------------------------
Insert Table 3 here
-----------------------------------

### 3.3.4 Supplementary Analysis—Exploring Interrelated Changes in Citation Behavior

The result that scientists who are more open are also the ones whose research output increased in innovativeness is consistent with these scientists being more inclined towards utilizing the new set of journals that they encountered because of AACR2. In Table 4, we investigate changes in citation behavior that might occur as a result of more open scientists citing a new set of journals following the AACR2 implementation.

-----------------------------------
Insert Table 4 here
-----------------------------------

In models M1 and M2, the dependent variable is the mean age of works cited in the focal publication. When scientists come across new journals, they are presumably likely to explore them by reading recent issues. Hence, we expect more open scientists to use more recently published articles following the AACR2 change. The coefficient of the interaction term between AACR2 and openness is -0.461 and significant ($p=0.000$). For the most open scientists, relative to the unconditional mean, this implies a 5.34% decrease in the age of the referenced articles used by those scientists. The finding suggests that more open scientists not only took up previously unused journals but also showed more of an inclination towards citing the relatively recent articles within those journals following the AACR2 change. The coefficient of the interaction term between AACR2 and depth of experience is insignificant ($p=0.397$).

      The second possible change following AACR2 is that open scientists could be expanding their collaborations. After being exposed to unfamiliar sources, scientists may be more inclined to collaborate with other scientists whose works are closer to those new journals. In Table 4 M3 and M4, the dependent variable is a network measure of team familiarity, with higher values representing publications by teams that have extensively worked together before on other publications and lower values representing teams that are working together for the first time. The coefficient of the interaction term between AACR2 and openness is -0.059 and significant ($p=0.016$). This shows that more open scientists are less likely to continue working with co-authors and more likely to find new co-authors after AACR2 is implemented.

Interestingly, the interaction term between AACR2 and depth of experience is 0.044 and significant ($p=0.008$), implying that scientists with a high depth of experience are more likely to work with familiar co-authors after AACR2 than scientists with a low depth of experience. One possibility is that as the works that scientists read were affected by AACR2—with more open scientists reading more widely and scientists with deep experience relying more on familiar publications—this impacted the collaborations that these scientists formed going forward.

## 4. ROBUSTNESS CHECKS

We ran additional tests to explore whether our results are robust to alternative measurement approaches and estimation techniques. In Table 5, we investigate the robustness of our findings regarding the effect of AACR2 on the innovativeness of publications using alternative measures of innovativeness: raw CD index of publications. In models M2 and M3, we include interaction terms for the implementation of AACR2 and the openness and depth of experience of scientists (i.e., corresponding authors of publications), respectively. In both models, the coefficients of AACR2 remain insignificant. In contrast, the coefficient on the interaction between AACR2 and openness is positive and significant in model M2 ($\beta = 0.026$, $p = 0.000$) and the interaction term between AACR2 and depth of experience is negative and significant in model M3 ($\beta = -0.013$, $p = 0.000$). Model M4, our preferred specification, includes both interaction terms. There, the coefficient on openness remains positive and significant ($\beta = 0.029$, $p = 0.000$), and the coefficient on the depth of experience is insignificant ($\beta = 0.005$, $p = 0.246$). Specifically, relative to the average CD index of all publications (0.086), scientists with the highest level of openness write papers with a CD index that are higher by 31.24% than scientists with the lowest level of openness following AACR2. Scientists with a high depth of experience in a particular field, however, do not appear to be impacted by AACR2. Thus, the results of this table using raw CD index are consistent with our analyses of the effect of AACR2 on the innovativeness of publications in Table 3.

-----------------------------------
Insert Table 5 here
-----------------------------------

In our main analyses, we assume that corresponding authors have influential roles in publication co-authoring teams so that their openness and depth of experience affect publication outcomes and citations made by publications. To provide support for our assumption, in Table 6, we exclude publications that are the results of co-authorships and rerun analyses with just single-author publications, which account for about 30.25% of publications. Results in Table 6 are generally consistent with our main analyses in Tables 2 and 3. Following AACR2 implementation, according to M2, very open scientists are more likely to produce destabilizing single-authored publications and less likely to produce consolidating single-authored publications (M2: $\beta = 0.049$, $p = 0.000$; M4: $\beta = -0.038$, $p = 0.000$). They are also more likely to cite a moved journal but less likely to cite a familiar journal post-AACR2 (M6: $\beta = 0.043$, $p = 0.000$; M10: $\beta = -0.066$, $p = 0.000$). On the other hand, publication innovativeness of scientists with a higher depth of experience is not likely to change following AACR2 for single-authored publications (M2: $\beta = -0.006$, $p = 0.603$; M2: $\beta = 0.006$, $p = 0.565$). As the results are generally consistent with our results in Table 2 and 3, Table 6 supports our assumption that corresponding authors lead the research process when co-working as a team with other scientists.

Another potential concern is that the results may only be generalizable to more productive scientists. Scientists that publish more are likely to be given more weight in publication-level analyses than scientists that publish fewer papers. To address this concern, we test the robustness of our findings with the scientist-panel in Table 7. The dependent variables are binary variables that indicate for each year whether the scientist publishes a destabilizing publication, publishes a consolidating publication, cites a moved journal, cites an unfamiliar journal, and cites a familiar journal. In M2, M4, M6, M8, and M10, we add corresponding author-fixed effects and use within-scientist variation to examine how changes in scientists' openness and depth of experience affect their publication outcome and citation behavior along with AACR2 implementation. Across all models in Table 7, results are generally consistent with our models in previous analyses. This provides further support to our findings that scientists' openness is critical for taking advantage of unexpected encounters.

In Table 8, we test openness with a different proxy variable: career age. Our rationale for using career age to proxy openness is that younger scientists have relatively fewer established collaborative partnerships to which they have to commit. When younger scientists come across new information, they have more cognitive bandwidth to explore the new information by collaborating with people outside of their department and field. Hence, we believe that younger scholars are more open than scientists that have been in the field for a longer time. We measured career age as the number of years that elapsed since the scientist's first publication year (Aschhoff and Grimpe 2014). Results in Table 8 are generally consistent with the results in Table 2 and 3. In M2, we find the interaction term between career age and AACR2 on *Destabilizing publication* to be negative and significant ($\beta = -0.0003$, $p = 0.002$), suggesting that younger scientists publish more destabilizing publications post-AACR2. The interaction term between career age and AACR2 on *Consolidating publication* in M4 is positive and significant ($\beta = 0.0002$, $p = 0.029$), which is consistent with results in Table 3 that show that more open scientists are less likely to produce consolidating publications. In M6 and M8, we also find that younger scientists are more likely to cite a moved journal ($\beta = -0.0004$, $p = 0.000$) and an unfamiliar journal following the AACR2 implementation ($\beta = -0.0002$, $p = 0.005$). The coefficients of the interaction term between the scientist's depth of experience and AACR2 are also consistent with results from Tables 2 and 3. The results of this table provide further support to our results from the main analyses that more open scientists are more likely to react to unexpected encounters while scientists with higher depth of experience are not.

-----------------------------------
Insert Tables 6, 7 and 8 here
-----------------------------------

## 5. DISCUSSION

In this paper, we leverage a unique natural experiment, the implementation of AACR2—in which the shelving locations of scientific journals were exogenously shifted across hundreds of North American research libraries—which allows us to closely examine serendipity in scientific discovery. Although prior studies note the important role that serendipity can play in scientific discoveries (Cannon 1940; Andel 1994), it has been difficult to study chance encounters in a structured manner given its nature. Therefore, it has been unclear which scientists tend to benefit from exposure to unsought information. Existing work points to the importance of a "prepared mind." However, scholars offer two different perspectives on what constitutes a prepared mind. On the one hand, some scholars emphasize the importance of an open mind, more willing to accept new ideas (Barber and Fox 1958). On the other hand, others argue that having deep experience in an area would be most helpful (Merton 1968; Roberts 1989).

Within the context of university libraries, we find that scientists that are more open are more likely to benefit from unplanned exposure to new information. Those with high openness produced more innovative work following AACR2 while scientists with high depth of experience produced less innovative work. These results reveal contrasting effects when the location of the journals that scientists use are shuffled and are potentially exposed to new sources of knowledge. Scientists that are more open are more likely to leverage exposure to new sources of information. In addition to utilizing journals that moved, these scientists specifically gravitate towards incorporating materials from journals they had not previously used. For scientists with a high depth of experience, creating friction in their access to these materials is challenging. When faced with this challenge, these scientists are less likely to publish innovative scientific works and more likely to draw upon familiar references. As Pasteur hypothesized, AACR2 and the exposure to serendipitously seen scientific material primarily benefited scientists whose previous work prepared them for incorporating novel information into their research.

Our findings provide general support for the view that a prepared mind is an open one. Open scientists filter through diverse types of unsought information they are exposed to and subsequently make new discoveries. Conversely, scientists with a deep depth of experience seem unable to deal effectively with new and unexpected information that does not neatly align with their preconceptions. Therefore, those with a range of knowledge that is broad seem better positioned to benefit from serendipitous encounters.

Although the study leverages a policy change in libraries, the focus on serendipity and discovery has implications for firm innovation. Firms looking to foster innovation may benefit from restructuring and reshuffling their organizations (e.g., Tushman, Newman, and Romanelli 1986; Romanelli & Tushman 1994; Zhang 2021) in ways that increase the chances of employees encountering unexpected information (March 1991; Tushman and O'Reilly 1996). Our study suggests that when individuals within the firm come across technological knowledge serendipitously, they may be able to utilize the knowledge to generate disruptive innovations. Nevertheless, there are likely to be both individual and organizational factors for managers to consider when implementing changes for serendipity. At the individual level, to benefit from restructuring, employees' cognitive frames must be receptive to unforeseen knowledge, which means that the employees' new ideas may be higher in novelty but lower in usefulness (Amabile 1983; Miron-Spketor and Beenen 2015). At the organizational level, such reorganizations are likely to deter exploitative capabilities which, depending on the state of the environment, may be suboptimal for the performance of the firm (March 1991, Teece, Pisano, and Shuen 1997). Therefore, our study suggests that managers should consider the nature of new technologies the firm needs and the dynamics of the competitive landscape when reorganizing for serendipity.

However, there are some limitations to our study. Our analysis is based on a policy change within physical libraries, which are less critical source of knowledge for research today than they once were. This may limit the generalizability of certain features of our study—for example, the application of the main identification strategy, a library policy change, to a more current context. However, we do believe that our main finding regarding the importance of a broader set of experiences has even greater implications for current scientists. Compared to before, today's scientists are likely exposed to even more unsought knowledge through recommendation systems of search engines and online research archives during the research process. Our findings suggest that scientists equipped with the ability to filter through seemingly unrelated information, using their broad knowledge, will benefit the most in today's research setting as well. In addition, in our study we are not able to accurately measure the actual physical distances that journals moved. The analysis is based on the assumption that the number of places a journal moves in alphabetical order roughly correlates to the amount of physical distance that a particular journal moved in the library. However, a journal moving a dozen places down the alphabetical order may simply mean the journal moves from the second to the third row of the same shelf. This implies that our measurement of relocation likely has some noise. However, we still find effects, suggesting that the actual effects of AACR2 may be even greater than what we report.

      Overall, in this paper we are able to examine the characteristics of scientists that facilitate serendipitous scientific discoveries. We believe this opens up promising directions for future research. A future study based in an online setting would likely provide ample opportunities to explore certain aspects of serendipity we were not able to explore, such as the potential importance of the frequency of exposure to unsought information. Scientists, even ones high in openness, may be willing to consider only a certain number of seemingly less-related information within a period of time before deciding to focus only on directly relevant knowledge. Similarly, a study that examines the role of serendipity in the context of management seems promising. In particular, research that examines which managers or firms are best positioned to take advantage of unforeseen opportunities would have interesting implications for the field of strategy (Levinthal and March 1993; Szulanski 1996; Dahlander, O'Mahony, and Gann 2016).

**FIGURE 1: CHANGE IN THE PROBABILITY OF CITING JOURNALS PROXIMATE TO AACR2 MOVED JOURNALS**

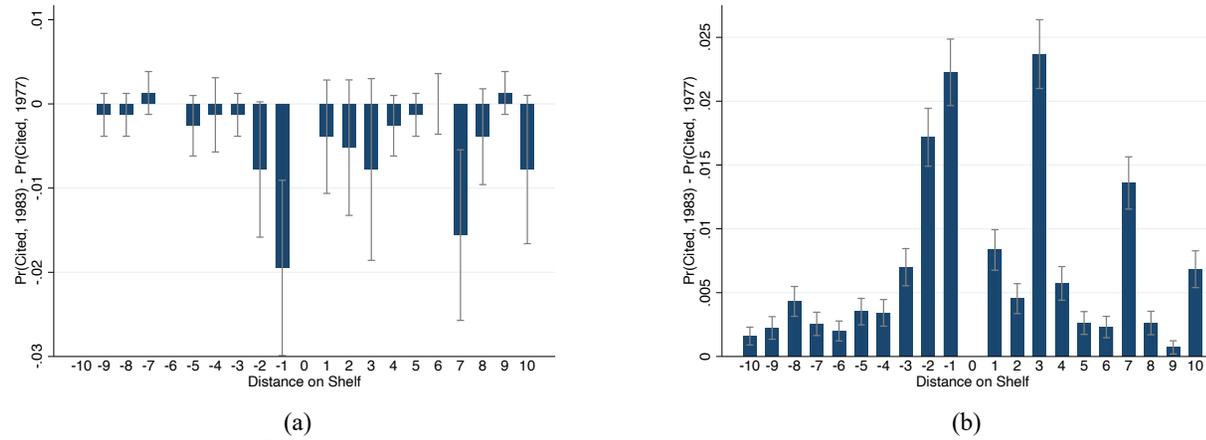

(a)          (b)

**Note:** The above plot shows the average change in probability of citing a journal between by a scientist in 1983 and 1977. For each journal that moved because of AACR2, we find the scientists who cited that journal in 1977. In plot (a), we show the average change in probability of citing that journal across the journals based on their shelf distance to the position that the moved journal used to occupy. In plot (b), we show the average change in probability of citing that journal across the journals based on their shelf distance to the position that the moved journal occupied after AACR2.

**FIGURE 2: EVENT STUDY OF AACR2'S EFFECT ON CITES TO JOURNALS**

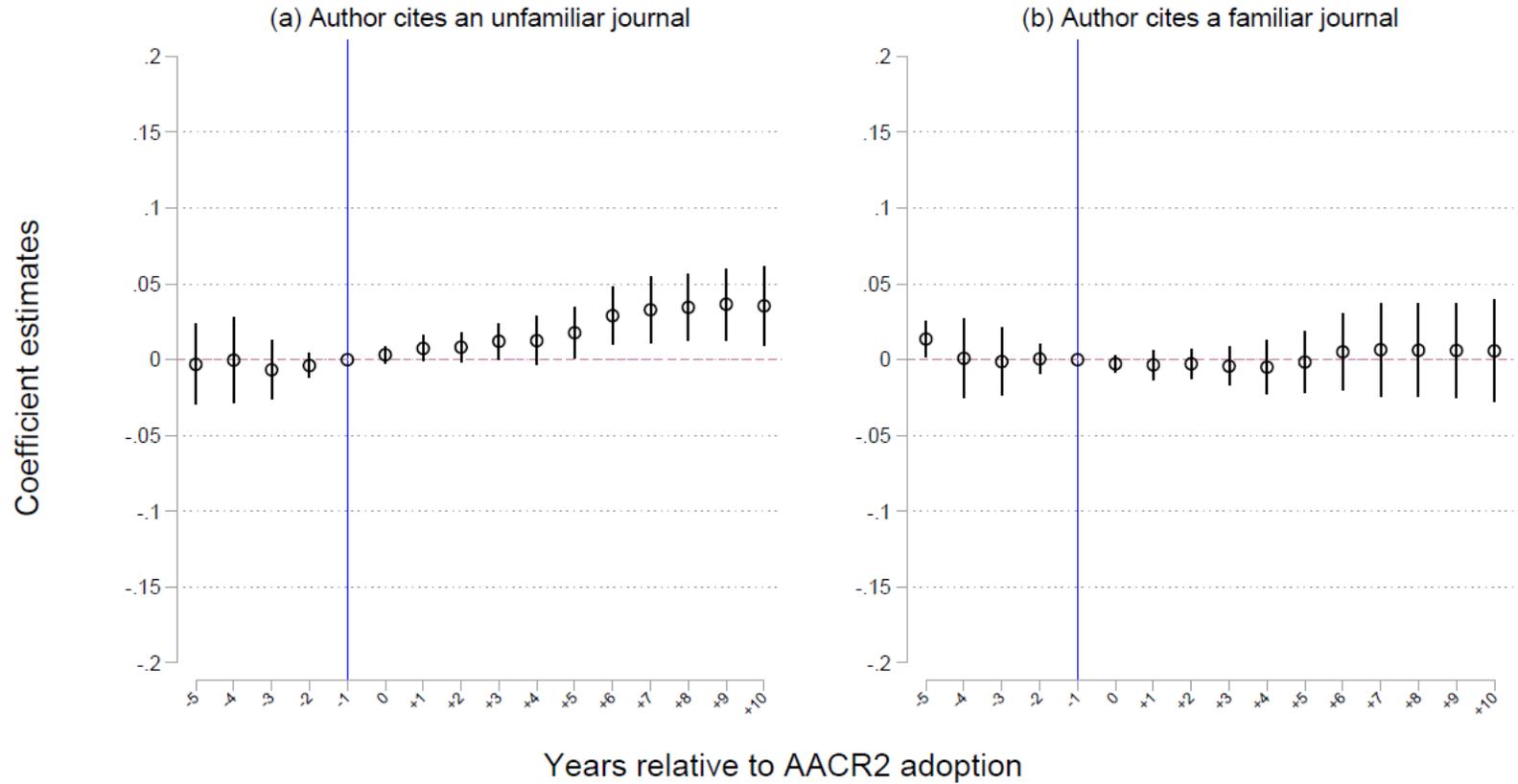

**Note:** The graphs above plot the changes in key dependent variables before and after the adoption of AACR2: (a) author's citation of an unfamiliar journal and (b) author's citation of a familiar journal. The data for the above plots is a panel in which an observation is a corresponding author x university in a year (some authors are affiliated to more than one university). The baseline year, marked by the blue line, is the year prior to the AACR2 adoption by the scientist's university. 95% confidence intervals. Robust standard errors clustered at the scientist level.

**TABLE 1: SUMMARY STATISTICS FOR VARIABLES USED IN REGRESSION ANALYSES**

| Variables | N | $N_{unique}$ | Mean | SD | Min | Max | Level of measurement |
|---|---|---|---|---|---|---|---|
| **Outcome** | | | | | | | |
| CD index | 2,281,786 | 329,970 | 0.101 | 0.296 | -1 | 1 | Publication |
| Cites a moved journal | 2,398,026 | 2 | 0.432 | 0.495 | 0 | 1 | Publication |
| Cites an unfamiliar journal | 2,398,026 | 2 | 0.693 | 0.461 | 0 | 1 | Publication |
| Cites a familiar journal | 2,398,026 | 2 | 0.674 | 0.469 | 0 | 1 | Publication |
| **Explanatory** | | | | | | | |
| Post-AACR2 | 2,398,026 | 2 | 0.876 | 0.329 | 0 | 1 | University x year |
| Openness (mean-centered) | 1,989,949 | 134,388 | -0.023 | 0.240 | -0.458 | 0.468 | Corresponding author x year |
| Depth of experience (mean-centered) | 2,120,523 | 22,046 | -0.004 | 0.249 | -0.671 | 0.313 | Corresponding author x year |
| **Controls** | | | | | | | |
| Ln(Number of publications) | 2,398,026 | 1,094 | 2.510 | 1.538 | 0 | 7.449 | Corresponding author x year |
| Ln(Number of references cited) | 2,398,026 | 293 | 2.210 | 1.152 | 0 | 7.621 | Publication |
| Ln(Publication team size) | 2,398,026 | 272 | 0.758 | 0.617 | 0 | 6.321 | Publication |
| **Fixed effects** | | | | | | | |
| Year | 2,398,026 | 20 | | | | | Year |
| Field | 2,398,026 | 151 | | | | | Field |
| University | 2,398,026 | 115 | | | | | University |

**TABLE 2: EFFECT OF AACR2 ON CITATION BEHAVIOR**

| | Cites a journal | | | | | |
|---|---|---|---|---|---|---|
| | Moved | | Unfamiliar (to the corresponding author) | | Familiar (to the corresponding author) | |
| | M1 | M2 | M3 | M4 | M5 | M6 |
| Openness (mean-centered) | 0.030*** | 0.003 | 0.124*** | 0.104*** | 0.059*** | 0.112*** |
| | (0.004) | (0.007) | (0.003) | (0.006) | (0.003) | (0.006) |
| Depth of experience (mean-centered) | -0.010* | -0.004 | -0.135*** | -0.121*** | 0.123*** | 0.104*** |
| | (0.004) | (0.007) | (0.003) | (0.006) | (0.003) | (0.005) |
| Post-AACR2 | 0.003 | 0.004 | 0.002 | 0.002 | -0.005 | -0.006* |
| | (0.004) | (0.005) | (0.003) | (0.003) | (0.003) | (0.003) |
| Post-AACR2 X Openness (mean-centered) | | 0.031*** | | 0.023*** | | -0.061*** |
| | | (0.008) | | (0.006) | | (0.006) |
| Post-AACR2 X Depth of experience (mean-centered) | | -0.006 | | -0.016* | | 0.022*** |
| | | (0.008) | | (0.006) | | (0.006) |
| Ln(Number of publications) | 0.001 | 0.001 | -0.108*** | -0.109*** | 0.039*** | 0.039*** |
| | (0.001) | (0.001) | (0.001) | (0.001) | (0.001) | (0.001) |
| Ln(Number of references cited) | 0.203*** | 0.203*** | 0.233*** | 0.233*** | 0.230*** | 0.230*** |
| | (0.001) | (0.001) | (0.001) | (0.001) | (0.001) | (0.001) |
| Ln(Publication team size) | 0.009*** | 0.009*** | 0.013*** | 0.013*** | 0.012*** | 0.012*** |
| | (0.002) | (0.002) | (0.001) | (0.001) | (0.001) | (0.001) |
| Field fixed effects | Yes | Yes | Yes | Yes | Yes | Yes |
| University fixed effects | Yes | Yes | Yes | Yes | Yes | Yes |
| Year fixed effects | Yes | Yes | Yes | Yes | Yes | Yes |
| N | 1,989,851 | 1,989,851 | 1,989,851 | 1,989,851 | 1,989,851 | 1,989,851 |
| Dep. Mean | 0.452 | 0.452 | 0.678 | 0.678 | 0.812 | 0.812 |
| R2 | 0.268 | 0.268 | 0.343 | 0.343 | 0.482 | 0.482 |

Note: These models show the relationship between the AACR2 implementation, corresponding author attributes, and citation behavior of corresponding authors. The dependent variables are whether or not the publication cites a moved journal, an unfamiliar journal, or a familiar journal. The results from M2 show that only the corresponding author's openness, and not the depth of experience, positively moderates the effect of AACR2 on the publication's likelihood of citing a moved journal. The results from M4 show that the corresponding author's openness positively moderates, while depth of experience negatively moderates the effect of AACR2 on the author's likelihood of citing an unfamiliar journal. The results M8 show that corresponding author's openness negatively moderates, while the depth of experience positively moderates, the effect of AACR2 on the author's likelihood of citing a familiar journal. Estimates are from linear probability models. Robust standard errors clustered at the university level in parentheses; *p*-values correspond to two-tailed tests.
*p<0.05, **p<0.01, ***p<0.001

**TABLE 3: EFFECT OF AACR2 ON PUBLICATION INNOVATIVENESS**

|  | Destabilizing publication | | Consolidating publication | |
| --- | --- | --- | --- | --- |
|  | M1 | M2 | M3 | M4 |
| Openness (mean-centered) | 0.049*** | 0.020** | -0.052*** | -0.032*** |
|  | (0.003) | (0.007) | (0.003) | (0.007) |
| Depth of experience (mean-centered) | -0.006* | -0.001 | 0.001 | -0.011 |
|  | (0.003) | (0.006) | (0.003) | (0.006) |
| Post-AACR2 | 0.005 | 0.005 | -0.009 | -0.009 |
|  | (0.005) | (0.005) | (0.005) | (0.005) |
| Post-AACR2 X Openness (mean-centered) |  | 0.033*** |  | -0.023** |
|  |  | (0.007) |  | (0.007) |
| Post-AACR2 X Depth of experience (mean-centered) |  | -0.006 |  | 0.013* |
|  |  | (0.007) |  | (0.006) |
| Ln(Number of publications) | 0.007*** | 0.007*** | -0.007*** | -0.007*** |
|  | (0.001) | (0.001) | (0.001) | (0.001) |
| Ln(Number of references cited) | -0.215*** | -0.215*** | 0.216*** | 0.216*** |
|  | (0.001) | (0.001) | (0.001) | (0.001) |
| Ln(Publication team size) | -0.016*** | -0.016*** | 0.046*** | 0.047*** |
|  | (0.002) | (0.002) | (0.002) | (0.002) |
| Field fixed effects | Yes | Yes | Yes | Yes |
| University fixed effects | Yes | Yes | Yes | Yes |
| Year fixed effects | Yes | Yes | Yes | Yes |
| N | 1,989,851 | 1,989,851 | 1,989,851 | 1,989,851 |
| Dep. Mean | 0.387 | 0.387 | 0.519 | 0.519 |
| R2 | 0.292 | 0.292 | 0.312 | 0.312 |

Note: These models show the relationship between AACR2, publication and corresponding author attributes, and measures for innovativeness of publications. The dependent variable is whether the publication is a destabilizing publication (CD index > 0) or not for models M1 and M2 and similarly, the dependent variable for models M3 and M4 is whether the publication is a consolidating publication (CD index < 0) or not. The results indicate that the corresponding author's openness positively moderates the effect of AACR2 on the likelihood of the publication being a destabilizing publication (M2), while it negatively moderates the likelihood of the publication being a consolidating publication (M4). Depth of experience only positively moderates the effect of AACR2 on the likelihood of the publication being a consolidating publication (M4). Estimates are from linear probability models. Robust standard errors clustered at the university level in parentheses; *p*-values correspond to two-tailed tests.
*p<0.05, **p<0.01, ***p<0.001

**TABLE 3: SUPPLEMENTARY ANALYSIS - EXPLORING INTERRELATED CHANGES**

|  | Mean age of work cited | | Team familiarity | |
|---|---|---|---|---|
|  | M1 | M2 | M3 | M4 |
| Openness (mean-centered) | 0.686*** | 1.090*** | -0.082*** | -0.029 |
|  | (0.063) | (0.119) | (0.012) | (0.023) |
| Depth of experience (mean-centered) | 0.570*** | 0.508*** | 0.114*** | 0.075*** |
|  | (0.049) | (0.083) | (0.010) | (0.017) |
| Post-AACR2 | -0.009 | -0.015 | 0.009 | 0.010 |
|  | (0.061) | (0.062) | (0.013) | (0.013) |
| Post-AACR2 X Openness (mean-centered) |  | -0.461*** |  | -0.059* |
|  |  | (0.114) |  | (0.024) |
| Post-AACR2 X Depth of experience (mean-centered) |  | 0.069 |  | 0.044** |
|  |  | (0.081) |  | (0.016) |
| Ln(Number of publications) | 0.008 | 0.009 | 0.136*** | 0.136*** |
|  | (0.011) | (0.011) | (0.003) | (0.003) |
| Ln(Number of references cited) | 0.084*** | 0.084*** | 0.096*** | 0.096*** |
|  | (0.018) | (0.018) | (0.002) | (0.002) |
| Ln(Publication team size) | -0.701*** | -0.700*** | 0.033*** | 0.033*** |
|  | (0.021) | (0.021) | (0.007) | (0.007) |
| Field fixed effects | Yes | Yes | Yes | Yes |
| University fixed effects | Yes | Yes | Yes | Yes |
| Year fixed effects | Yes | Yes | Yes | Yes |
| N | 1,781,916 | 1,781,916 | 1,410,177 | 1,410,177 |
| Dep. Mean | 8.565 | 8.565 | 0.809 | 0.809 |
| R2 | 0.057 | 0.057 | 0.120 | 0.120 |

Note: These models show the relationship between AACR2 implementation, corresponding author attributes, and other publication characteristics. The dependent variables are the mean age of works cited in the publication in M1 and M2, and a measure of co-author familiarity in M3 and M4. The corresponding author's openness negatively moderates the effect of AACR2 on the mean age work cited in the publication (M2) and familiarity of the publication team (M4). An author's depth of experience only positively moderates the effect of AACR2 on the author's likelihood of citing work familiar to the team (M4). Estimates are from ordinary least squares regression models. Robust standard errors clustered at the university level in parentheses; $p$-values correspond to two-tailed tests.
*p<0.05, **p<0.01, ***p<0.001

**TABLE 5: ROBUSTNESS CHECK—ALTERNATIVE MEASUREMENT OF INNOVATIVENESS OF PUBLICATIONS**

|  | CD Index | | | |
|---|---|---|---|---|
|  | M1 | M2 | M3 | M4 |
| Openness (mean-centered) | -0.004* | -0.027*** | -0.004* | -0.029*** |
|  | (0.002) | (0.003) | (0.002) | (0.004) |
| Depth of experience (mean-centered) | -0.018*** | -0.018*** | -0.007** | -0.023*** |
|  | (0.001) | (0.001) | (0.002) | (0.004) |
| Post-AACR2 | -0.000 | 0.000 | -0.000 | 0.000 |
|  | (0.003) | (0.003) | (0.003) | (0.003) |
| Post-AACR2 X Openness (mean-centered) |  | 0.026*** |  | 0.029*** |
|  |  | (0.003) |  | (0.004) |
| Post-AACR2 X Depth of experience (mean-centered) |  |  | -0.013*** | 0.005 |
|  |  |  | (0.003) | (0.004) |
| Ln(Number of publications) | 0.003*** | 0.003*** | 0.003*** | 0.003*** |
|  | (0.000) | (0.000) | (0.000) | (0.000) |
| Ln(Number of references cited) | -0.196*** | -0.196*** | -0.196*** | -0.196*** |
|  | (0.001) | (0.001) | (0.001) | (0.001) |
| Ln(Publication team size) | -0.005*** | -0.005*** | -0.005*** | -0.005*** |
|  | (0.001) | (0.001) | (0.001) | (0.001) |
| Field fixed effects | Yes | Yes | Yes | Yes |
| University fixed effects | Yes | Yes | Yes | Yes |
| Year fixed effects | Yes | Yes | Yes | Yes |
| N | 1,918,088 | 1,918,088 | 1,918,088 | 1,918,088 |
| Dep. Mean | 0.086 | 0.086 | 0.086 | 0.086 |
| R2 | 0.513 | 0.513 | 0.513 | 0.513 |

Note: These models show the relationship between AACR2 implementation, corresponding author attributes, and innovativeness of publications. The dependent variable is the CD index of each publication. The results from M2 and M4 suggest that the corresponding author's openness has a significant positive moderation effect on publication innovativeness. Results from M3 show that the corresponding author's depth of experience has a negative and significant moderation effect on publication innovativeness, but in M4, the author's depth of experience does not have a significant moderation effect on publication innovativeness. These results are consistent with our main results in Table 3. Estimates are from ordinary least squares regression models. Robust standard errors clustered at the university level in parentheses; *p*-values correspond to two-tailed tests.
*p<0.05, **p<0.01, ***p<0.001

**TABLE 4: ROBUSTNESS CHECK—ANALYSES WITH SINGLE AUTHOR PUBLICATIONS**

| | Destabilizing publication | | Consolidating publication | | Cites a moved journal | | Cites an unfamiliar journal (to the corresponding author) | | Cites a familiar journal (to the corresponding author) | |
|---|---|---|---|---|---|---|---|---|---|---|
| | M1 | M2 | M3 | M4 | M5 | M6 | M7 | M8 | M9 | M10 |
| Openness (mean-centered) | 0.038*** | -0.004 | -0.043*** | -0.011 | 0.035*** | -0.002 | 0.098*** | 0.096*** | 0.048*** | 0.104*** |
| | (0.005) | (0.011) | (0.004) | (0.009) | (0.005) | (0.009) | (0.005) | (0.009) | (0.004) | (0.008) |
| Depth of experience (mean-centered) | -0.017*** | -0.012 | 0.005 | -0.000 | -0.008 | -0.001 | -0.121*** | -0.109*** | 0.129*** | 0.126*** |
| | (0.004) | (0.010) | (0.004) | (0.009) | (0.005) | (0.009) | (0.004) | (0.009) | (0.004) | (0.009) |
| Post-AACR2 | 0.005 | 0.007 | -0.006 | -0.008 | 0.006 | 0.009 | -0.001 | -0.001 | -0.003 | -0.007 |
| | (0.010) | (0.010) | (0.008) | (0.008) | (0.006) | (0.006) | (0.006) | (0.006) | (0.005) | (0.005) |
| Post-AACR2 X Openness (mean-centered) | | 0.049*** | | -0.038*** | | 0.043*** | | 0.002 | | -0.066*** |
| | | (0.011) | | (0.010) | | (0.010) | | (0.009) | | (0.010) |
| Post-AACR2 X Depth of experience (mean-centered) | | -0.006 | | 0.006 | | -0.007 | | -0.014 | | 0.003 |
| | | (0.011) | | (0.010) | | (0.010) | | (0.010) | | (0.010) |
| Ln(Number of publications) | 0.012*** | 0.011*** | -0.012*** | -0.012*** | 0.003*** | 0.003*** | -0.101*** | -0.101*** | 0.038*** | 0.038*** |
| | (0.001) | (0.001) | (0.001) | (0.001) | (0.001) | (0.001) | (0.001) | (0.001) | (0.001) | (0.001) |
| Ln(Number of references cited) | -0.203*** | -0.202*** | 0.188*** | 0.188*** | 0.189*** | 0.190*** | 0.251*** | 0.251*** | 0.262*** | 0.262*** |
| | (0.001) | (0.001) | (0.001) | (0.001) | (0.001) | (0.001) | (0.001) | (0.001) | (0.001) | (0.001) |
| Field fixed effects | Yes | Yes | Yes | Yes | Yes | Yes | Yes | Yes | Yes | Yes |
| University fixed effects | Yes | Yes | Yes | Yes | Yes | Yes | Yes | Yes | Yes | Yes |
| Year fixed effects | Yes | Yes | Yes | Yes | Yes | Yes | Yes | Yes | Yes | Yes |
| N | 579,674 | 579,674 | 579,674 | 579,674 | 579,674 | 579,674 | 579,674 | 579,674 | 579,674 | 579,674 |
| Dep. Mean | 0.516 | 0.516 | 0.356 | 0.356 | 0.333 | 0.333 | 0.577 | 0.577 | 0.689 | 0.689 |
| R2 | 0.318 | 0.318 | 0.326 | 0.326 | 0.320 | 0.320 | 0.390 | 0.390 | 0.569 | 0.569 |

Note: To address the possibility that the corresponding author may not lead the research process, in this table, we replicate Tables 2 and 3 with single-authored publications. All coefficients of interaction terms are similar with those in Tables 2 and 3, suggesting that corresponding authors do have considerable influence on the research process. Estimates are from linear probability models. Robust standard errors clustered at the university level in parentheses; *p*-values correspond to two-tailed tests.
*p<0.05, **p<0.01, ***p<0.001

**TABLE 7: ROBUSTNESS CHECK—AUTHOR-LEVEL ANALYSES OF EFFECT OF AACR2**

| | Publishes a destabilizing paper | | Publishes a consolidating paper | | Cites a moved journal | | Cites an unfamiliar journal | | Cites a familiar journal | |
|---|---|---|---|---|---|---|---|---|---|---|
| | M1 | M2 | M3 | M4 | M5 | M6 | M7 | M8 | M9 | M10 |
| Openness (mean-centered) | -0.072*** | 0.001 | -0.033*** | 0.017* | 0.028*** | 0.006 | 0.028*** | 0.005 | 0.042*** | 0.204*** |
| | (0.009) | (0.008) | (0.006) | (0.007) | (0.006) | (0.006) | (0.005) | (0.006) | (0.005) | (0.006) |
| Depth of experience (mean-centered) | 0.028** | 0.029** | 0.001 | -0.003 | -0.010 | 0.002 | -0.054*** | -0.034*** | 0.097*** | -0.003 |
| | (0.009) | (0.010) | (0.005) | (0.006) | (0.005) | (0.006) | (0.005) | (0.006) | (0.005) | (0.006) |
| Post-AACR2 | 0.007* | 0.003 | -0.007* | -0.006 | -0.000 | 0.001 | 0.001 | 0.000 | -0.002 | -0.004** |
| | (0.003) | (0.004) | (0.003) | (0.004) | (0.002) | (0.003) | (0.002) | (0.002) | (0.002) | (0.001) |
| Post-AACR2 X Openness (mean-centered) | 0.041*** | 0.053*** | -0.011* | 0.006 | 0.007 | -0.010 | 0.017*** | 0.013* | -0.047*** | -0.013* |
| | (0.008) | (0.008) | (0.006) | (0.006) | (0.005) | (0.006) | (0.004) | (0.006) | (0.005) | (0.005) |
| Post-AACR2 X Depth of experience (mean-centered) | 0.006 | -0.022* | 0.001 | 0.001 | 0.004 | -0.001 | -0.014** | -0.020*** | -0.013** | -0.027*** |
| | (0.007) | (0.009) | (0.005) | (0.006) | (0.005) | (0.006) | (0.005) | (0.005) | (0.005) | (0.006) |
| Ln(Number of publications) | 0.097*** | 0.114*** | 0.051*** | 0.058*** | 0.027*** | 0.014*** | 0.026*** | 0.028*** | 0.065*** | 0.081*** |
| | (0.005) | (0.005) | (0.002) | (0.002) | (0.002) | (0.001) | (0.002) | (0.001) | (0.001) | (0.001) |
| Ln(Total number of references cited (year)) | 0.020*** | 0.053*** | 0.218*** | 0.206*** | 0.191*** | 0.194*** | 0.249*** | 0.263*** | 0.265*** | 0.271*** |
| | (0.001) | (0.001) | (0.001) | (0.001) | (0.001) | (0.001) | (0.001) | (0.001) | (0.001) | (0.001) |
| Ln(Average publication team size (year)) | 0.008*** | 0.075*** | -0.002*** | -0.017*** | 0.003*** | 0.003* | -0.067*** | -0.060*** | 0.017*** | 0.055*** |
| | (0.001) | (0.002) | (0.001) | (0.001) | (0.000) | (0.001) | (0.001) | (0.002) | (0.000) | (0.001) |
| University fixed effects | Yes | Yes | Yes | Yes | Yes | Yes | Yes | Yes | Yes | Yes |
| Author fixed effects | No | Yes | No | Yes | No | Yes | No | Yes | No | Yes |
| Year fixed effects | Yes | Yes | Yes | Yes | Yes | Yes | Yes | Yes | Yes | Yes |
| N | 1,634,973 | 1,466,110 | 1,634,973 | 1,466,110 | 1,634,973 | 1,466,110 | 1,634,973 | 1,466,110 | 1,634,973 | 1,466,110 |
| Dep. Mean | 0.275 | 0.258 | 0.347 | 0.329 | 0.302 | 0.285 | 0.457 | 0.418 | 0.502 | 0.481 |
| R2 | 0.048 | 0.327 | 0.557 | 0.670 | 0.446 | 0.626 | 0.646 | 0.722 | 0.762 | 0.850 |

Note: In this table, we replicate Tables 2 and 3 with the scientist-panel. All publication-level variables are aggregated to the author-year level. In Models M2, M4, M6, M8, and M10, we include author fixed effects to investigate the within-author effects. Estimates are from linear probability models. Robust standard errors clustered at the university level in parentheses; *p*-values correspond to two-tailed tests.
*p<0.05, **p<0.01, ***p<0.001

**TABLE 8: ROBUSTNESS CHECK— SCIENTIST'S CAREER AGE AS A PROXY FOR OPENNESS**

| | Destabilizing publication | | Consolidating publication | | Cites a moved journal | | Cites an unfamiliar journal (to the corresponding author) | | Cites a familiar journal (to the corresponding author) | |
|---|---|---|---|---|---|---|---|---|---|---|
| | M1 | M2 | M3 | M4 | M5 | M6 | M7 | M8 | M9 | M10 |
| Openness (mean-centered) | 0.000 | 0.000*** | -0.000*** | -0.000*** | -0.000*** | 0.000 | 0.002*** | 0.002*** | -0.003*** | -0.003*** |
| | (0.000) | (0.000) | (0.000) | (0.000) | (0.000) | (0.000) | (0.000) | (0.000) | (0.000) | (0.000) |
| Depth of experience (mean-centered) | -0.026*** | 0.004 | 0.021*** | -0.004 | -0.023*** | 0.008 | -0.167*** | -0.151*** | -0.010*** | -0.075*** |
| | (0.002) | (0.005) | (0.002) | (0.004) | (0.004) | (0.006) | (0.003) | (0.005) | (0.003) | (0.005) |
| Post-AACR2 | 0.004 | 0.009 | -0.009 | -0.012* | 0.002 | 0.008 | 0.002 | 0.006 | -0.005 | -0.006 |
| | (0.005) | (0.005) | (0.005) | (0.006) | (0.004) | (0.005) | (0.003) | (0.003) | (0.003) | (0.003) |
| Post-AACR2 X Openness (mean-centered) | | -0.000** | | 0.000* | | -0.000*** | | -0.000** | | 0.000 |
| | | (0.000) | | (0.000) | | (0.000) | | (0.000) | | (0.000) |
| Post-AACR2 X Depth of experience (mean-centered) | | -0.034*** | | 0.028*** | | -0.036*** | | -0.018*** | | 0.075*** |
| | | (0.005) | | (0.005) | | (0.006) | | (0.005) | | (0.004) |
| Ln(Number of publications) | 0.005*** | 0.005*** | -0.005*** | -0.004*** | 0.002** | 0.002** | -0.121*** | -0.121*** | 0.086*** | 0.086*** |
| | (0.001) | (0.001) | (0.001) | (0.001) | (0.001) | (0.001) | (0.001) | (0.001) | (0.001) | (0.001) |
| Ln(Number of references cited) | -0.216*** | -0.216*** | 0.215*** | 0.215*** | 0.201*** | 0.201*** | 0.241*** | 0.241*** | 0.224*** | 0.224*** |
| | (0.001) | (0.001) | (0.001) | (0.001) | (0.001) | (0.001) | (0.001) | (0.001) | (0.001) | (0.001) |
| Ln(Publication team size) | -0.016*** | -0.016*** | 0.046*** | 0.046*** | 0.010*** | 0.010*** | 0.015*** | 0.015*** | 0.009*** | 0.009*** |
| | (0.002) | (0.002) | (0.002) | (0.002) | (0.001) | (0.001) | (0.001) | (0.001) | (0.001) | (0.001) |
| Field fixed effects | Yes | Yes | Yes | Yes | Yes | Yes | Yes | Yes | Yes | Yes |
| University fixed effects | Yes | Yes | Yes | Yes | Yes | Yes | Yes | Yes | Yes | Yes |
| Year fixed effects | Yes | Yes | Yes | Yes | Yes | Yes | Yes | Yes | Yes | Yes |
| N | 2,120,523 | 2,120,523 | 2,120,523 | 2,120,523 | 2,120,523 | 2,120,523 | 2,120,523 | 2,120,523 | 2,120,523 | 2,120,523 |
| Dep. Mean | 0.403 | 0.403 | 0.503 | 0.503 | 0.441 | 0.441 | 0.676 | 0.676 | 0.762 | 0.762 |
| R2 | 0.309 | 0.310 | 0.328 | 0.328 | 0.279 | 0.279 | 0.362 | 0.362 | 0.490 | 0.491 |

Note: In this table, we proxy corresponding author's openness with career age (i.e., years since PhD attainment) and replicate Tables 2 and 3. All coefficients of interaction terms are mostly similar with those in Tables 2 and 3, suggesting that our measure of openness may be a good proxy for openness. Estimates are from linear probability models. Robust standard errors clustered at the university level in parentheses; *p*-values correspond to two-tailed tests.
*p<0.05, **p<0.01, ***p<0.001

**APPENDIX A: EXAMINING DIFFERENCES BETWEEN JOURNALS WITH OR WITHOUT CORPORATE AUTHOR'S NAME**

|  | Journals with Corporate Author's Name | Journals without Corporate Author's Name | Two-tailed p-value |
|---|---|---|---|
| Average Number of Citations Received per Year Prior to AACR2 | 598.685 | 504.046 | 0.204 |

Notes: We compare whether the inclusion of corporate author name in the journal's title make scientists more likely to cite the journal. We calculate the average number of citations that journals receive per year before the AACR2 implementation. We then compare whether the average citations received are different by whether journals have corporate author's name included or not. Results show that the average citations of two types of journals are not statistically different ($p$ = 0.204). This provides evidence that the inclusion of corporate author name does not make scientists more likely to cite the journal.

**APPENDIX B: DATA ON CORPORATE AUTHOR NAMES OF JOURNALS**

**Data Construction and Coding**
After merging the WoS and Library of Congress Catalog data, we searched the MARC 21 records for all journals to identify those with corporate authors, after which we extracted the name of the corporate author.[13] Subsequently, journals were manually flagged as "moved" by two members of the team if the journal title included the name of the corporate author. As an initial step, two members from the team first went through a subset of journals independently to identify journals that actually included the corporate authors' names in the titles of journals (e.g., Journal of American Medical Association). Discrepancies between the two were resolved by a third member of the team. Based on the flagging results from the subset, two members independently went through the complete list and each identified journal titles with corporate authors.

**Coding Procedure**
In this section, we provide five illustrative examples of journal title names, corporate names, and our manual coding of the treatment from our raw data. Then to demonstrate our coding procedure, we explain how e coded the treatment across these different journals, title names, and corporate names.

|   | Journal title name | Corporate names | AACR2 treated |
|---|---|---|---|
| A | 'Academy of Management Journal' | 'Academy of Management', 'AOM' | 1 |
| B | 'Monthly Weather Review' | 'American Meteorological Society.', 'United States.' | 0 |
| C | 'Mind' | 'Oxford University Press.', 'Mind Association.' | 0 |
| D | 'British Medical Journal' | 'British Medical Association.' | 0 |
| E | 'Clinical toxicology : the official journal of the American Academy of Clinical Toxicology and European Association of Poisons Centres and Clinical Toxicologists.' | 'American Academy of Clinical Toxicology.', 'European Association of Poisons Centres and Clinical Toxicologists.', 'American Association of Poison Control Centers.' | 1 |

- **Case A**—The name of the journal ('Academy of Management Journal') includes the full name of the corporate body ('Academy of Management') so is coded as being treated by AACR2.

- **Case B**—The name of the journal ('Monthly Weather Review') does not include the full name of the corporate body ('American Meteorological Society') so is coded as not being treated by AACR2.

- **Case C**—The name of the journal is 'Mind' and the publisher is 'Oxford University Press' on behalf of the 'Mind Association'. Full name of either publisher is not included in the journal name, so is coded as not being treated by AACR2.
    - As in this case and Case E, multiple different corporate bodies could be provided as publishers of one single journal name. In these cases, it is coded as being treated by AACR2 if one corporate body's full name appears in the journal's name.

- **Case D**—The name of the journal ('British Medical Journal') partially included the corporate body ('British Medical Association'). However, because there is one word from the corporate body that is not included in the journal title, it is being not treated by AACR2. Full corporate names need to be included in the journal title in order for it to be considered as treated by AACR2.

- **Case E**—There are three different names for corporate bodies ('American Academy of Clinical Toxicology', 'European Association of Poisons Centres and Clinical Toxicologists', 'American Association of Poison Control

---

[13] Using MARC fields 110, 610, 710, and 810.

Centers') and two of them are included in the longer version of the journal's name. Therefore, it is coded as being treated by AACR2.
- o As in this case and Case C, multiple different corporate bodies could be provided as publishers of one single journal name. In these cases, it is coded as being treated by AACR2 if one corporate body's full name appears in the journal's name.

**APPENDIX C: ANALYSES EXCLUDING UNIVERSITY LIBRARIES WITH IMPUTED AACR2 IMPLEMENTATION YEARS**

| | Destabilizing publication | | Consolidating publication | | Cites a moved journal | | Cites an unfamiliar journal | | Cites a familiar journal | |
|---|---|---|---|---|---|---|---|---|---|---|
| | M1 | M2 | M3 | M4 | M5 | M6 | M7 | M8 | M9 | M10 |
| Openness (mean-centered) | 0.047*** | 0.016 | -0.047*** | -0.025* | 0.032*** | 0.007 | 0.122*** | 0.100*** | 0.062*** | 0.119*** |
| | (0.004) | (0.010) | (0.004) | (0.010) | (0.007) | (0.009) | (0.005) | (0.008) | (0.004) | (0.008) |
| Depth of experience (mean-centered) | -0.006 | 0.006 | 0.000 | -0.015 | -0.005 | 0.009 | -0.136*** | -0.110*** | 0.124*** | 0.106*** |
| | (0.004) | (0.009) | (0.003) | (0.008) | (0.006) | (0.010) | (0.005) | (0.010) | (0.006) | (0.008) |
| Post-AACR2 | 0.007 | 0.007 | -0.011 | -0.011* | 0.005 | 0.005 | 0.000 | 0.000 | -0.008** | -0.009** |
| | (0.005) | (0.005) | (0.005) | (0.005) | (0.005) | (0.005) | (0.003) | (0.003) | (0.003) | (0.003) |
| Post-AACR2 X Openness (mean-centered) | | 0.035*** | | -0.026* | | 0.029* | | 0.025** | | -0.066*** |
| | | (0.010) | | (0.011) | | (0.011) | | (0.008) | | (0.010) |
| Post-AACR2 X Depth of experience (mean-centered) | | -0.013 | | 0.017 | | -0.016 | | -0.029** | | 0.020* |
| | | (0.009) | | (0.009) | | (0.013) | | (0.009) | | (0.009) |
| Ln(Number of publications) | 0.007*** | 0.007*** | -0.008*** | -0.007*** | 0.001 | 0.001 | -0.108*** | -0.109*** | 0.039*** | 0.039*** |
| | (0.001) | (0.001) | (0.001) | (0.001) | (0.001) | (0.001) | (0.001) | (0.001) | (0.001) | (0.001) |
| Ln(Number of references cited) | -0.214*** | -0.214*** | 0.216*** | 0.216*** | 0.201*** | 0.201*** | 0.232*** | 0.232*** | 0.230*** | 0.230*** |
| | (0.001) | (0.001) | (0.001) | (0.001) | (0.002) | (0.002) | (0.001) | (0.001) | (0.002) | (0.002) |
| Ln(Publication team size) | -0.017*** | -0.017*** | 0.045*** | 0.046*** | 0.008** | 0.008** | 0.012*** | 0.012*** | 0.013*** | 0.013*** |
| | (0.003) | (0.003) | (0.003) | (0.003) | (0.003) | (0.003) | (0.001) | (0.001) | (0.001) | (0.001) |
| Field fixed effects | Yes | Yes | Yes | Yes | Yes | Yes | Yes | Yes | Yes | Yes |
| University fixed effects | No | Yes | No | Yes | No | Yes | No | Yes | No | Yes |
| Year fixed effects | Yes | Yes | Yes | Yes | Yes | Yes | Yes | Yes | Yes | Yes |
| N | 887,105 | 887,105 | 887,105 | 887,105 | 887,105 | 887,105 | 887,105 | 887,105 | 887,105 | 887,105 |
| Dep. Mean | 0.388 | 0.388 | 0.520 | 0.520 | 0.451 | 0.451 | 0.678 | 0.678 | 0.813 | 0.813 |
| R2 | 0.292 | 0.292 | 0.314 | 0.314 | 0.267 | 0.267 | 0.343 | 0.343 | 0.479 | 0.479 |

Note: In this table, we exclude university libraries with 1981 as the imputed year of AACR2 implementation and replicate Tables 2 and 3. All coefficients of interaction terms are mostly similar with those in Tables 2 and 3. These results add confidence that the subset of universities with imputed AACR2 implementation years do not systematically change the relationships that we observe in our analyses. Estimates are from linear probability models. Robust standard errors clustered at the university level in parentheses; *p*-values correspond to two-tailed tests.
*p<0.05, **p<0.01, ***p<0.001

**APPENDIX D: CROSS TAB OF SCIENTISTS' OPENNESS AND DEPTH OF EXPERIENCE**

|  | Low Depth of Experience | High Depth of Experience | Total |
|---|---|---|---|
| Low Openness | 233,930 | 583,736 | 817,666 |
| High Openness | 651,993 | 542,319 | 1,402,258 |
| Total | 885,923 | 1,126,055 | 2,011,978 |

**Note:** The table above includes all observations at the author-year level. Openness and experience depth are split by the median value. This table shows that all four quadrants of openness and depth of experience have significant number of observations, suggesting that scientists can have high (low) openness and low (high) depth of experience. Observations are not concentrated in one or two quadrants.

**APPENDIX E: CORRELATIONS OF KEY VARIABLES USED IN REGRESSION ANALYSES**

|      | Variable | (1) | (2) | (3) | (4) | (5) | (6) | (7) | (8) | (9) | (10) |
|------|----------|-----|-----|-----|-----|-----|-----|-----|-----|-----|------|
| (1)  | Post-AACR2 | 1.000 | | | | | | | | | |
| (2)  | CD index | -0.029 | 1.000 | | | | | | | | |
| (3)  | Cites a moved journal | 0.043 | -0.285 | 1.000 | | | | | | | |
| (4)  | Cites an unfamiliar journal | 0.015 | -0.414 | 0.205 | 1.000 | | | | | | |
| (5)  | Cites a familiar journal | 0.036 | -0.663 | 0.263 | 0.163 | 1.000 | | | | | |
| (6)  | Openness (mean-centered) | 0.010 | -0.017 | 0.080 | 0.066 | 0.071 | 1.000 | | | | |
| (7)  | Depth of experience (mean-centered) | -0.006 | 0.009 | -0.067 | -0.028 | -0.039 | -0.665 | 1.000 | | | |
| (8)  | Ln(Number of publications) | 0.059 | -0.016 | 0.057 | -0.246 | 0.160 | 0.302 | -0.405 | 1.000 | | |
| (9)  | Ln(Number of references cited) | 0.072 | -0.691 | 0.427 | 0.437 | 0.611 | 0.121 | -0.100 | 0.074 | 1.000 | |
| (10) | Ln(Publication team size) | 0.062 | -0.146 | 0.138 | 0.086 | 0.155 | 0.117 | -0.078 | 0.039 | 0.234 | 1.000 |

# APPENDIX F: EFFECT OF AACR2 ON CITATION BEHAVIOR USING COUNTS OF JOURNALS CITED

|  | DV: Number of moved journals cited | | DV: Number of unfamiliar journals cited | | Number of familiar journals cited | |
|---|---|---|---|---|---|---|
|  | M1 | M2 | M3 | M4 | M5 | M6 |
| Openness (mean-centered) | -0.219*** | -0.253*** | 0.357*** | 0.218*** | 0.252*** | 0.315*** |
|  | (0.022) | (0.041) | (0.011) | (0.015) | (0.009) | (0.017) |
| Depth of experience (mean-centered) | 0.078*** | 0.160*** | -0.409*** | -0.368*** | 0.127*** | 0.047*** |
|  | (0.020) | (0.031) | (0.008) | (0.013) | (0.008) | (0.014) |
| Post-AACR2 | -0.002 | -0.004 | 0.014 | 0.011 | -0.009 | -0.001 |
|  | (0.019) | (0.019) | (0.010) | (0.009) | (0.010) | (0.010) |
| Post-AACR2 X Openness (mean-centered) |  | 0.038 |  | 0.156*** |  | -0.070*** |
|  |  | (0.037) |  | (0.015) |  | (0.015) |
| Post-AACR2 X Depth of experience (mean-centered) |  | -0.091** |  | -0.045** |  | 0.089*** |
|  |  | (0.031) |  | (0.014) |  | (0.013) |
| Ln(Number of publications) | 0.036*** | 0.036*** | -0.417*** | -0.417*** | 0.164*** | 0.165*** |
|  | (0.003) | (0.003) | (0.002) | (0.002) | (0.001) | (0.001) |
| Ln(Number of references cited) | 1.028*** | 1.027*** | 0.852*** | 0.852*** | 0.832*** | 0.832*** |
|  | (0.005) | (0.005) | (0.002) | (0.002) | (0.003) | (0.003) |
| Ln(Publication team size) | 0.023*** | 0.023*** | 0.054*** | 0.054*** | 0.024*** | 0.024*** |
|  | (0.005) | (0.005) | (0.002) | (0.002) | (0.001) | (0.001) |
| Field fixed effects | Yes | Yes | Yes | Yes | Yes | Yes |
| University fixed effects | Yes | Yes | Yes | Yes | Yes | Yes |
| Year fixed effects | Yes | Yes | Yes | Yes | Yes | Yes |
| N | 1,989,851 | 1,989,851 | 1,989,851 | 1,989,851 | 1,989,851 | 1,989,851 |
| Dep. Mean | 1.369 | 1.369 | 2.574 | 2.574 | 5.076 | 5.076 |
| Log Likelihood | -3,065,054.612 | -3064952.740 | -3780639.810 | -3780190.887 | -4018356.074 | -4017846.079 |
| Chi Squared | 45,865.720 | 49,476.884 | 236,799.609 | 244,550.432 | 96,899.591 | 97,024.076 |

Note: In this table, we replicate Tables 2 by using the counts of journals cited in each publication as dependent variables. All coefficients of interaction terms are mostly similar with those in Table 2. Estimates are from pseudo-maximum likelihood Poisson models. Robust standard errors clustered at the university level in parentheses; *p*-values correspond to two-tailed tests.

**APPENDIX G: NUMBER OF UNIVERSITIES AND OBSERVATIONS BY ADOPTION YEARS**

| AACR2 adoption years | Without imputed adoption years | | With imputed adoption years | |
|---|---|---|---|---|
| | Number of universities | Number of observations | Number of universities | Number of observations |
| **1979** | 2 | 55,738 | 2 | 55,738 |
| **1980** | 7 | 264,612 | 7 | 264,612 |
| **1981** | 31 | 655,406 | 101 | 1,985,230 |
| **1982** | 2 | 17,825 | 2 | 17,825 |
| **1984** | 1 | 42,578 | 1 | 42,578 |
| **1988** | 1 | 7,390 | 1 | 7,390 |
| **1989** | 1 | 24,653 | 1 | 24,653 |
| **Total** | 45 | 1,068,202 | 115 | 2,398,026 |

Note: In this table, we report the number of university libraries and the number of observations by AACR2 adoption years of university libraries.

# APPENDIX H: ANALYSES WITHOUT EARLY ADOPTERS AS CONTROLS

|  | Destabilizing publication | | Consolidating publication | | Cites a moved journal | | Cites an unfamiliar journal (to the corresponding author) | | Cites a familiar journal (to the corresponding author) | |
|---|---|---|---|---|---|---|---|---|---|---|
|  | M1 | M2 | M3 | M4 | M5 | M6 | M7 | M8 | M9 | M10 |
| Openness (mean-centered) | 0.052*** | 0.026 | -0.053*** | -0.028 | 0.030* | 0.021 | 0.106*** | 0.087*** | 0.097*** | 0.116*** |
|  | (0.011) | (0.018) | (0.008) | (0.013) | (0.011) | (0.011) | (0.011) | (0.014) | (0.005) | (0.008) |
| Depth of experience (mean-centered) | 0.027 | 0.003 | -0.029** | -0.002 | 0.004 | -0.004 | -0.121*** | -0.132*** | 0.108*** | 0.102*** |
|  | (0.012) | (0.013) | (0.007) | (0.009) | (0.014) | (0.023) | (0.011) | (0.016) | (0.009) | (0.011) |
| Post-AACR2 | 0.011 | 0.013 | -0.017* | -0.019* | 0.010 | 0.011* | 0.002 | 0.003 | -0.018* | -0.018* |
|  | (0.007) | (0.007) | (0.006) | (0.006) | (0.005) | (0.004) | (0.009) | (0.009) | (0.006) | (0.006) |
| Post-AACR2 X Openness (mean-centered) |  | 0.048* |  | -0.046* |  | 0.016 |  | 0.035* |  | -0.033 |
|  |  | (0.019) |  | (0.016) |  | (0.026) |  | (0.014) |  | (0.015) |
| Post-AACR2 X Depth of experience (mean-centered) |  | 0.044 |  | -0.050* |  | 0.014 |  | 0.021 |  | 0.012 |
|  |  | (0.020) |  | (0.021) |  | (0.026) |  | (0.014) |  | (0.014) |
| Ln(Number of publications) | 0.009*** | 0.009*** | -0.010** | -0.010** | 0.002 | 0.002 | -0.112*** | -0.112*** | 0.039*** | 0.039*** |
|  | (0.002) | (0.002) | (0.002) | (0.002) | (0.006) | (0.005) | (0.002) | (0.002) | (0.003) | (0.003) |
| Ln(Number of references cited) | -0.211*** | -0.211*** | 0.212*** | 0.212*** | 0.198*** | 0.198*** | 0.239*** | 0.239*** | 0.236*** | 0.236*** |
|  | (0.003) | (0.003) | (0.004) | (0.004) | (0.004) | (0.004) | (0.004) | (0.004) | (0.006) | (0.006) |
| Ln(Publication team size) | -0.020** | -0.020** | 0.048*** | 0.048*** | 0.003 | 0.003 | 0.009* | 0.009* | 0.017** | 0.017** |
|  | (0.004) | (0.004) | (0.003) | (0.003) | (0.006) | (0.006) | (0.004) | (0.004) | (0.004) | (0.004) |
| Field fixed effects | Yes | Yes | Yes | Yes | Yes | Yes | Yes | Yes | Yes | Yes |
| University fixed effects | Yes | Yes | Yes | Yes | Yes | Yes | Yes | Yes | Yes | Yes |
| Year fixed effects | Yes | Yes | Yes | Yes | Yes | Yes | Yes | Yes | Yes | Yes |
| N | 89,532 | 89,532 | 89,532 | 89,532 | 89,532 | 89,532 | 89,532 | 89,532 | 89,532 | 89,532 |
| Dep. Mean | 0.403 | 0.403 | 0.506 | 0.506 | 0.424 | 0.424 | 0.672 | 0.672 | 0.802 | 0.802 |
| R2 | 0.291 | 0.291 | 0.315 | 0.315 | 0.260 | 0.260 | 0.357 | 0.357 | 0.486 | 0.486 |

Note: In this table, we limit the sample to 1980, 1984, 1988, and 1989 AACR2 adopting universities to address the concern that early adopters become controls for later adopters. We limit observations up to year 1983 for universities adopting in 1980 and up to year 1987 for universities adopting in 1984 We replicate Tables 2 and 3. All coefficients of interaction terms are mostly similar with those in Tables 2 and 3. Estimates are from linear probability models. Robust standard errors clustered at the university level in parentheses; *p*-values correspond to two-tailed tests.
*p<0.05, **p<0.01, ***p<0.001